\documentclass[aps,prd,twocolumn,superscriptaddress,showpacs]{revtex4}

\usepackage{graphicx}
\usepackage{dcolumn}
\usepackage{bm}
\usepackage{lineno}
\usepackage{setspace}
\usepackage{subfig}
\usepackage{booktabs}
\usepackage{rotating}
\usepackage{hyperref}
\usepackage{float}
\begin{document}

\newcommand\et{E_T}
\newcommand\Ht{H_T}
\newcommand\pt{p_T}
\newcommand\ttbar{t\bar{t}}
\newcommand\ppbar{p\bar{p}}
\newcommand\bbbar{b\bar{b}}
\newcommand\ccbar{c\bar{c}}



\title{Measurement of the $\ttbar$ Production Cross Section in $\ppbar$ Collisions at $\sqrt{s}$=1.96 TeV using Soft Electron $b$-Tagging}
\affiliation{Institute of Physics, Academia Sinica, Taipei, Taiwan 11529, Republic of China} 
\affiliation{Argonne National Laboratory, Argonne, Illinois 60439} 
\affiliation{University of Athens, 157 71 Athens, Greece} 
\affiliation{Institut de Fisica d'Altes Energies, Universitat Autonoma de Barcelona, E-08193, Bellaterra (Barcelona), Spain} 
\affiliation{Baylor University, Waco, Texas  76798} 
\affiliation{Istituto Nazionale di Fisica Nucleare Bologna, $^y$University of Bologna, I-40127 Bologna, Italy} 
\affiliation{Brandeis University, Waltham, Massachusetts 02254} 
\affiliation{University of California, Davis, Davis, California  95616} 
\affiliation{University of California, Los Angeles, Los Angeles, California  90024} 
\affiliation{University of California, San Diego, La Jolla, California  92093} 
\affiliation{University of California, Santa Barbara, Santa Barbara, California 93106} 
\affiliation{Instituto de Fisica de Cantabria, CSIC-University of Cantabria, 39005 Santander, Spain} 
\affiliation{Carnegie Mellon University, Pittsburgh, PA  15213} 
\affiliation{Enrico Fermi Institute, University of Chicago, Chicago, Illinois 60637}
\affiliation{Comenius University, 842 48 Bratislava, Slovakia; Institute of Experimental Physics, 040 01 Kosice, Slovakia} 
\affiliation{Joint Institute for Nuclear Research, RU-141980 Dubna, Russia} 
\affiliation{Duke University, Durham, North Carolina  27708} 
\affiliation{Fermi National Accelerator Laboratory, Batavia, Illinois 60510} 
\affiliation{University of Florida, Gainesville, Florida  32611} 
\affiliation{Laboratori Nazionali di Frascati, Istituto Nazionale di Fisica Nucleare, I-00044 Frascati, Italy} 
\affiliation{University of Geneva, CH-1211 Geneva 4, Switzerland} 
\affiliation{Glasgow University, Glasgow G12 8QQ, United Kingdom} 
\affiliation{Harvard University, Cambridge, Massachusetts 02138} 
\affiliation{Division of High Energy Physics, Department of Physics, University of Helsinki and Helsinki Institute of Physics, FIN-00014, Helsinki, Finland} 
\affiliation{University of Illinois, Urbana, Illinois 61801} 
\affiliation{The Johns Hopkins University, Baltimore, Maryland 21218} 
\affiliation{Institut f\"{u}r Experimentelle Kernphysik, Universit\"{a}t Karlsruhe, 76128 Karlsruhe, Germany} 
\affiliation{Center for High Energy Physics: Kyungpook National University, Daegu 702-701, Korea; Seoul National University, Seoul 151-742, Korea; Sungkyunkwan University, Suwon 440-746, Korea; Korea Institute of Science and Technology Information, Daejeon 305-806, Korea; Chonnam National University, Gwangju 500-757, Korea; Chonbuk National University, Jeonju 561-756, Korea} 
\affiliation{Ernest Orlando Lawrence Berkeley National Laboratory, Berkeley, California 94720} 
\affiliation{University of Liverpool, Liverpool L69 7ZE, United Kingdom} 
\affiliation{University College London, London WC1E 6BT, United Kingdom} 
\affiliation{Centro de Investigaciones Energeticas Medioambientales y Tecnologicas, E-28040 Madrid, Spain} 
\affiliation{Massachusetts Institute of Technology, Cambridge, Massachusetts  02139} 
\affiliation{Institute of Particle Physics: McGill University, Montr\'{e}al, Qu\'{e}bec, Canada H3A~2T8; Simon Fraser University, Burnaby, British Columbia, Canada V5A~1S6; University of Toronto, Toronto, Ontario, Canada M5S~1A7; and TRIUMF, Vancouver, British Columbia, Canada V6T~2A3} 
\affiliation{University of Michigan, Ann Arbor, Michigan 48109} 
\affiliation{Michigan State University, East Lansing, Michigan  48824}
\affiliation{Institution for Theoretical and Experimental Physics, ITEP, Moscow 117259, Russia} 
\affiliation{University of New Mexico, Albuquerque, New Mexico 87131} 
\affiliation{Northwestern University, Evanston, Illinois  60208} 
\affiliation{The Ohio State University, Columbus, Ohio  43210} 
\affiliation{Okayama University, Okayama 700-8530, Japan} 
\affiliation{Osaka City University, Osaka 588, Japan} 
\affiliation{University of Oxford, Oxford OX1 3RH, United Kingdom} 
\affiliation{Istituto Nazionale di Fisica Nucleare, Sezione di Padova-Trento, $^z$University of Padova, I-35131 Padova, Italy} 
\affiliation{LPNHE, Universite Pierre et Marie Curie/IN2P3-CNRS, UMR7585, Paris, F-75252 France} 
\affiliation{University of Pennsylvania, Philadelphia, Pennsylvania 19104}
\affiliation{Istituto Nazionale di Fisica Nucleare Pisa, $^{aa}$University of Pisa, $^{bb}$University of Siena and $^{cc}$Scuola Normale Superiore, I-56127 Pisa, Italy} 
\affiliation{University of Pittsburgh, Pittsburgh, Pennsylvania 15260} 
\affiliation{Purdue University, West Lafayette, Indiana 47907} 
\affiliation{University of Rochester, Rochester, New York 14627} 
\affiliation{The Rockefeller University, New York, New York 10021} 
\affiliation{Istituto Nazionale di Fisica Nucleare, Sezione di Roma 1, $^{dd}$Sapienza Universit\`{a} di Roma, I-00185 Roma, Italy} 

\affiliation{Rutgers University, Piscataway, New Jersey 08855} 
\affiliation{Texas A\&M University, College Station, Texas 77843} 
\affiliation{Istituto Nazionale di Fisica Nucleare Trieste/Udine, I-34100 Trieste, $^{ee}$University of Trieste/Udine, I-33100 Udine, Italy} 
\affiliation{University of Tsukuba, Tsukuba, Ibaraki 305, Japan} 
\affiliation{Tufts University, Medford, Massachusetts 02155} 
\affiliation{Waseda University, Tokyo 169, Japan} 
\affiliation{Wayne State University, Detroit, Michigan  48201} 
\affiliation{University of Wisconsin, Madison, Wisconsin 53706} 
\affiliation{Yale University, New Haven, Connecticut 06520} 
\author{T.~Aaltonen}
\affiliation{Division of High Energy Physics, Department of Physics, University of Helsinki and Helsinki Institute of Physics, FIN-00014, Helsinki, Finland}
\author{J.~Adelman}
\affiliation{Enrico Fermi Institute, University of Chicago, Chicago, Illinois 60637}
\author{T.~Akimoto}
\affiliation{University of Tsukuba, Tsukuba, Ibaraki 305, Japan}
\author{B.~\'{A}lvarez~Gonz\'{a}lez$^t$}
\affiliation{Instituto de Fisica de Cantabria, CSIC-University of Cantabria, 39005 Santander, Spain}
\author{S.~Amerio$^z$}
\affiliation{Istituto Nazionale di Fisica Nucleare, Sezione di Padova-Trento, $^z$University of Padova, I-35131 Padova, Italy} 

\author{D.~Amidei}
\affiliation{University of Michigan, Ann Arbor, Michigan 48109}
\author{A.~Anastassov}
\affiliation{Northwestern University, Evanston, Illinois  60208}
\author{A.~Annovi}
\affiliation{Laboratori Nazionali di Frascati, Istituto Nazionale di Fisica Nucleare, I-00044 Frascati, Italy}
\author{J.~Antos}
\affiliation{Comenius University, 842 48 Bratislava, Slovakia; Institute of Experimental Physics, 040 01 Kosice, Slovakia}
\author{G.~Apollinari}
\affiliation{Fermi National Accelerator Laboratory, Batavia, Illinois 60510}
% ...

\collaboration{CDF Collaboration\footnote{With visitors from $^a$University of Massachusetts Amherst, Amherst, Massachusetts 01003,
$^b$Universiteit Antwerpen, B-2610 Antwerp, Belgium, 
$^c$University of Bristol, Bristol BS8 1TL, United Kingdom,
$^d$Chinese Academy of Sciences, Beijing 100864, China, 
$^e$Istituto Nazionale di Fisica Nucleare, Sezione di Cagliari, 09042 Monserrato (Cagliari), Italy,
$^f$University of California Irvine, Irvine, CA  92697, 
$^g$University of California Santa Cruz, Santa Cruz, CA  95064, 
$^h$Cornell University, Ithaca, NY  14853, 
$^i$University of Cyprus, Nicosia CY-1678, Cyprus, 
$^j$University College Dublin, Dublin 4, Ireland,
$^k$University of Edinburgh, Edinburgh EH9 3JZ, United Kingdom, 
$^l$University of Fukui, Fukui City, Fukui Prefecture, Japan 910-0017
$^m$Kinki University, Higashi-Osaka City, Japan 577-8502
$^n$Universidad Iberoamericana, Mexico D.F., Mexico,
$^o$University of Iowa, Iowa City, IA  52242,
$^p$Queen Mary, University of London, London, E1 4NS, England,
$^q$University of Manchester, Manchester M13 9PL, England, 
$^r$Nagasaki Institute of Applied Science, Nagasaki, Japan, 
$^s$University of Notre Dame, Notre Dame, IN 46556,
$^t$University de Oviedo, E-33007 Oviedo, Spain, 
$^u$Texas Tech University, Lubbock, TX  79609, 
$^v$IFIC(CSIC-Universitat de Valencia), 46071 Valencia, Spain,
$^w$University of Virginia, Charlottesville, VA  22904,
$^x$Bergische Universit\"at Wuppertal, 42097 Wuppertal, Germany,
$^{ff}$On leave from J.~Stefan Institute, Ljubljana, Slovenia, 
}}
\noaffiliation


\vspace*{2.0cm}

\begin{abstract}
We present a measurement of the top quark pair production cross section in $\ppbar$ collisions at $\sqrt{s}$=1.96 TeV using a data sample corresponding to 1.7 fb$^{-1}$ of integrated luminosity collected with the Collider Detector at Fermilab.  We reconstruct $\ttbar$ events in the lepton+jets channel, consisting of $e\nu$+jets and $\mu\nu$+jets final states.  The dominant background is the production of $W$ bosons in association with multiple jets.  To suppress this background, we identify electrons from the semileptonic decay of heavy-flavor jets (`soft electron tags').  From a sample of 2196 candidate events, we obtain 120 tagged events with a background expectation of $51\pm3$ events, corresponding to a cross section of $\sigma_{\ttbar}=7.8\pm2.4$\,(stat)\,$\pm\,1.6$\,(syst)\,$\pm\,0.5$\,(lumi) pb.  We assume a top-quark mass of 175~GeV/$c^2$.  This is the first measurement of the $\ttbar$ cross section with soft electron tags in Run II of the Tevatron.
\end{abstract}


\maketitle


\section{\label{sec:Intro}Introduction}

The top quark is the most massive fundamental particle observed to date, and has been studied by the CDF and D0 collaborations since its discovery in 1995 \cite{TopDiscovery}.  The $\ttbar$ production cross section has been measured in each of the three canonical final states: $q\bar{q}' b\, q\bar{q}'\bar{b}$ \cite{TopHadronic}, $q\bar{q}'b\, \ell\bar{\nu} \bar{b}$ \cite{TopKinematic, TopSecVtx, TopSLTmu}, and $\bar{\ell}\nu b\, \ell\bar{\nu} \bar{b}$ \cite{TopDilepton} ($\ell=e$, $\mu$, and $q=u$, $d$, $c$, $s$).  In these measurements, different combinations of $b$-quark identification (`tagging') and kinematic information \cite{TopKinematic} have been used to suppress backgrounds.  Tagging of $b$-quarks has been accomplished by identifying the long lifetime of the hadron with secondary vertex reconstruction or with displaced tracks \cite{TopSecVtx} or through soft muons from semileptonic decay \cite{TopSLTmu}.  Along with measurements of the top-quark mass \cite{TopMass} and many other properties of the top quark, a consistent picture of the top quark as the third generation standard model (SM) isospin partner of the bottom quark emerges.

The Fermilab Tevatron produces top quarks, typically in pairs, by colliding $\ppbar$ at $\sqrt{s}=1.96$~TeV.  The $\ttbar$ production cross section calculated at next-to-leading order is $6.7\pm0.8$ pb \cite{TopXS} assuming $m_t=175$~GeV/$c^2$, where the uncertainty is dominated by the choice of renormalization and factorization scales.  At the Tevatron, approximately 85\% of $\ttbar$ production is via quark-antiquark annihilation and 15\% is via gluon-gluon fusion.  The measurement of the production cross section is important first as a test of perturbative QCD, but also as a platform from which to study other top-quark properties.  Moreover, measuring the $\ttbar$ cross section in its various final states is an important consistency test of the SM and might highlight contributions to a particular decay channel from new physics.

In this paper, we present a measurement of the $\ttbar$ production cross section in the lepton plus $\ge3$~jets final state.  The dominant background in this channel is the production of a $W$ boson associated with several jets.  To suppress this background, we use a soft electron tagger (SLT$_\mathrm{e}$) to identify the semileptonic decay of heavy-flavor (HF).  Heavy-flavor refers to the product of the fragmentation of a bottom or charm quark.

Soft electron tagging is a challenging method of identifying $b$-jets because the semileptonic branching fraction (BF) is approximately 20\% $-$ BF($b\to e\nu X$) and BF($b\to c\to e\nu X$) each contribute approximately 10\% $-$ and because electron identification is complicated by the presence of a surrounding jet.  The algorithm is able to distinguish electromagnetic showers from hadronic showers by using a shower-maximum detector embedded in the electromagnetic calorimeter.  This detector has a high enough resolution that it can determine the transverse shape and position of electron showers and yet be unaffected by nearby activity.  Additionally, $\gamma\to e^+ e^-$ conversions due to material interactions provide a significant background, which we suppress using a combination of geometric and kinematic requirements.  Nevertheless, the soft electron technique is interesting because it is complementary to other $b$-tagging techniques and because it is a useful technique for other analyses.

This is the first measurement of the $\ttbar$ cross section with soft electron tags in Run II of the Tevatron.  A previous measurement at $\sqrt{s}=1.8$~TeV combined secondary vertex tagging, soft muon, and soft electron tagging~\cite{TopRunI}.

We organize this paper as follows: Sec.~\ref{sec:cdf} describes aspects of the CDF detector salient to this analysis.  Section~\ref{sec:slte} describes the implementation of the SLT$_\mathrm{e}$.  We discuss the SLT$_\mathrm{e}$ tagging efficiency in $\ttbar$ events in Sec.~\ref{sec:efficiency}.  Section~\ref{sec:fakes} describes the calculation of the background to tagged electrons in HF jets, including conversion electrons and hadrons.  In Sec.~\ref{sec:bbbar}, we tune the SLT$_\mathrm{e}$ tagger in a $\bbbar$ control sample.  This ensures the tagger's validity in high-momentum $b$-jets, such as those found in $\ttbar$ events.  Section~\ref{sec:xs} reports the cross section measurement, including the event selection and signal and background estimation.  Finally, in Sec.~\ref{sec:conclusions} we present our results and conclusions.

\section{The CDF Detector}\label{sec:cdf}

\begin{figure*}[htb]
\begin{center}
\includegraphics[width=6in]{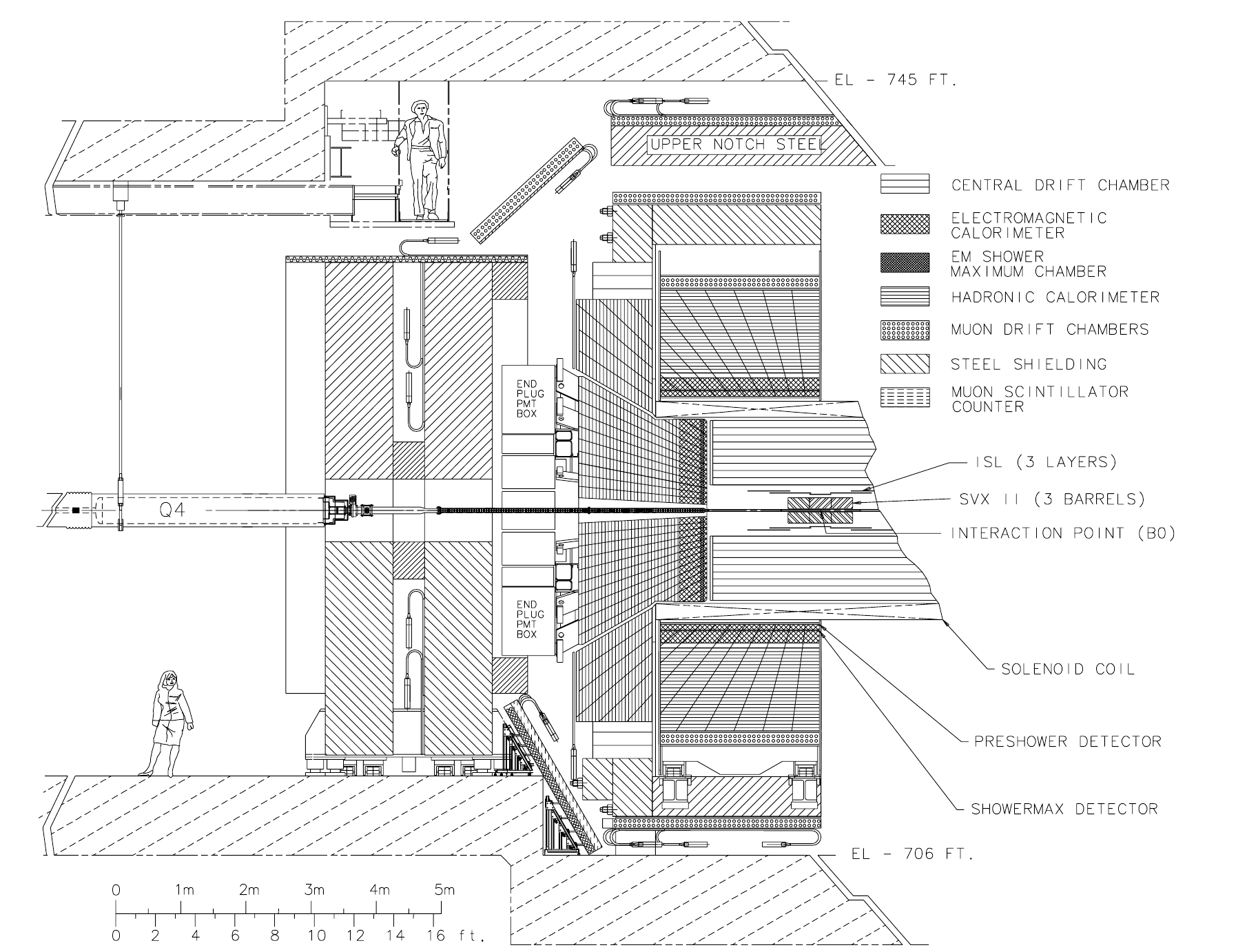}
\caption{Illustration of the CDF II detector.} \label{fig:cdf}
\end{center}
\end{figure*}

CDF II is a multi-purpose, azimuthally and forward-backward symmetric detector designed to study $\ppbar$ collisions at the Tevatron.  An illustration of the detector is shown in Fig.~\ref{fig:cdf}.  We use a cylindrical coordinate system where $z$ points along the proton direction, $\phi$ is the azimuthal angle about the beam axis, and $\theta$ is the polar angle to the proton beam direction.  We define the pseudorapidity $\eta\equiv-\ln\tan(\theta/2)$.

The tracking system consists of silicon microstrip detectors and an open-cell drift chamber immersed in a 1.4 T solenoidal magnetic field.  The silicon microstrip detectors provide precise charged particle tracking in the radial range from 1.5 $-$ 28 cm.  The silicon detectors are divided into three different subcomponents, comprised of eight total layers.  Layer00 (L00) \cite{L00} is a single-sided silicon detector mounted directly on the beampipe.  The silicon vertex detector (SVXII) \cite{SVXII} consists of five double-sided sensors with radial range up to 10.6 cm.  The intermediate silicon layer (ISL) \cite{ISL} is composed of two layers of double-sided silicon, extending coverage up to $|\eta|<2.0$.  The drift chamber, referred to as the central outer tracker (COT) \cite{COT}, consists of 96 layers of sense wires grouped in 8 alternating superlayers of axial and stereo wires, covering a radial range from 40 to 140 cm.  The reconstructed trajectories of COT tracks are extrapolated into the silicon detectors, and the track is refit using the additional hits in the silicon detectors.  In combination, the COT and silicon detectors provide excellent tracking up to $|\eta|\le 1.1$.  The transverse momentum ($\pt$) resolution, $\sigma(\pt)/\pt$, is approximately $0.07\%\, \pt$ [GeV/$c$]$^{-1}$ when hits from the SVXII and ISL are included.

Beyond the solenoid lie the electromagnetic and hadronic calorimeters, with coverage up to $|\eta|\le 3.6$.  The calorimeters have a projective geometry with a segmentation of $\Delta\eta\approx0.1$ and $\Delta\phi\approx15^\circ$ in the central ($|\eta|\le 1.1$) region.  The central electromagnetic calorimeter (CEM) \cite{CEM} consists of $>18$ radiation lengths ($X_0$) of lead-scintillator sandwich and contains wire and strip chambers embedded at the expected shower maximum ($\sim 6X_0$).  The wire and strip chambers are collectively referred to as the central shower-maximum (CES) chambers and provide measurements of the transverse electromagnetic shower shape along the $r-\phi$ and $z$ directions with a resolution of 1 and 2 mm, respectively.  The central hadronic calorimeter (CHA) \cite{CHA} consists of $\sim$\,4.7 interaction lengths of alternating lead-scintillator layers at normal incidence.  Measured in units of GeV, the CEM has an energy resolution, $\sigma(E)/E=13.5\%/\sqrt{E\sin(\theta)}\oplus 2\%$, and the CHA has an energy resolution, $\sigma(E)/E=50\%/\sqrt{E}$.

Muon chambers \cite{Muons} consist of layers of drift tubes surrounding the calorimeter.  The central muon detector (CMU) is cylindrical and covers a pseudorapidity range $|\eta|<0.63$.  The central muon upgrade (CMP) is a box-shaped set of drift chambers located beyond the CMU and separated by more than three interaction lengths of steel.  Muons which produce hits in both the CMU and CMP are called CMUP.  The central muon extension (CMX) extends the muon coverage up to $|\eta|\le 1$.

Gaseous Cherenkov luminosity counters (CLC) \cite{CLC} provide the luminosity measurement with a $\pm\, 6\%$ relative uncertainty.

CDF uses a three-level trigger system to select events to be recorded to tape.  The first two levels perform a limited set of reconstruction with dedicated hardware, and the third level is a software trigger performing speed-optimized event reconstruction algorithms.  The triggers used in this analysis include electron, muon, and jet triggers at different transverse energy thresholds.  The electron triggers require the coincidence of a track with an electromagnetic cluster in the central calorimeter.  The muon triggers require a track that points to hits in the muon chambers.  The jet triggers require calorimeter clusters with uncorrected $\et$ above a specified threshold.

\section{Soft Electron Tagging}\label{sec:slte}

The SLT$_\mathrm{e}$ algorithm uses the COT and silicon trackers, central calorimeter and, in particular, the central shower maximum chambers to identify electrons embedded in jets from semileptonic decays of HF quarks.  The tagging algorithm is `track-based' - as opposed to `jet-based' - in that we consider every track in the event that meets certain criteria as a candidate for tagging.  Such tracks are required to be well-measured by the COT and to extrapolate to the CES.  This requirement forces the track to have $|\eta|$ less than 1.2.  We require that the track $\pt$ is greater than 2~GeV/$c$.  We consider only tracks that originate close to the primary vertex: $|d_0|<0.3$~cm, $|z_0|<60$~cm, and $|z_0-z_{\mathrm{vtx}}|<5$~cm, where $d_0$ is the impact parameter, that is the distance of closest approach in the transverse plane, with respect to the beamline.  The $z$ position of the track at closest approach to the beamline is $z_0$, and $z_{\mathrm{vtx}}$ is the reconstructed $z$ position of the primary vertex.  Tracks must also pass a jet-matching requirement, which is that they are within $\Delta R\equiv\sqrt{\Delta\eta^2+\Delta\phi^2}\le0.4$ from the axis of a jet with transverse energy $\et$ greater than 20~GeV.  Jets are clustered with a fixed-cone algorithm with a cone of size $\Delta R\le0.4$.  Jet energies are corrected for detector response, multiple interactions, and un-instrumented regions of the detector \cite{JES}.  Finally, tracks must also pass a conversion filter described in Sec.~\ref{sec:conversions}.  Although we have not explicitly required tracks to have silicon hits, the conversion filter insists that tracks with a high number of `missing' silicon hits must be discarded.  We consider tracks which meet all of the above criteria as SLT$_\mathrm{e}$ candidates.

\begin{figure}[h]
\begin{center}
\includegraphics[width=3.0in]{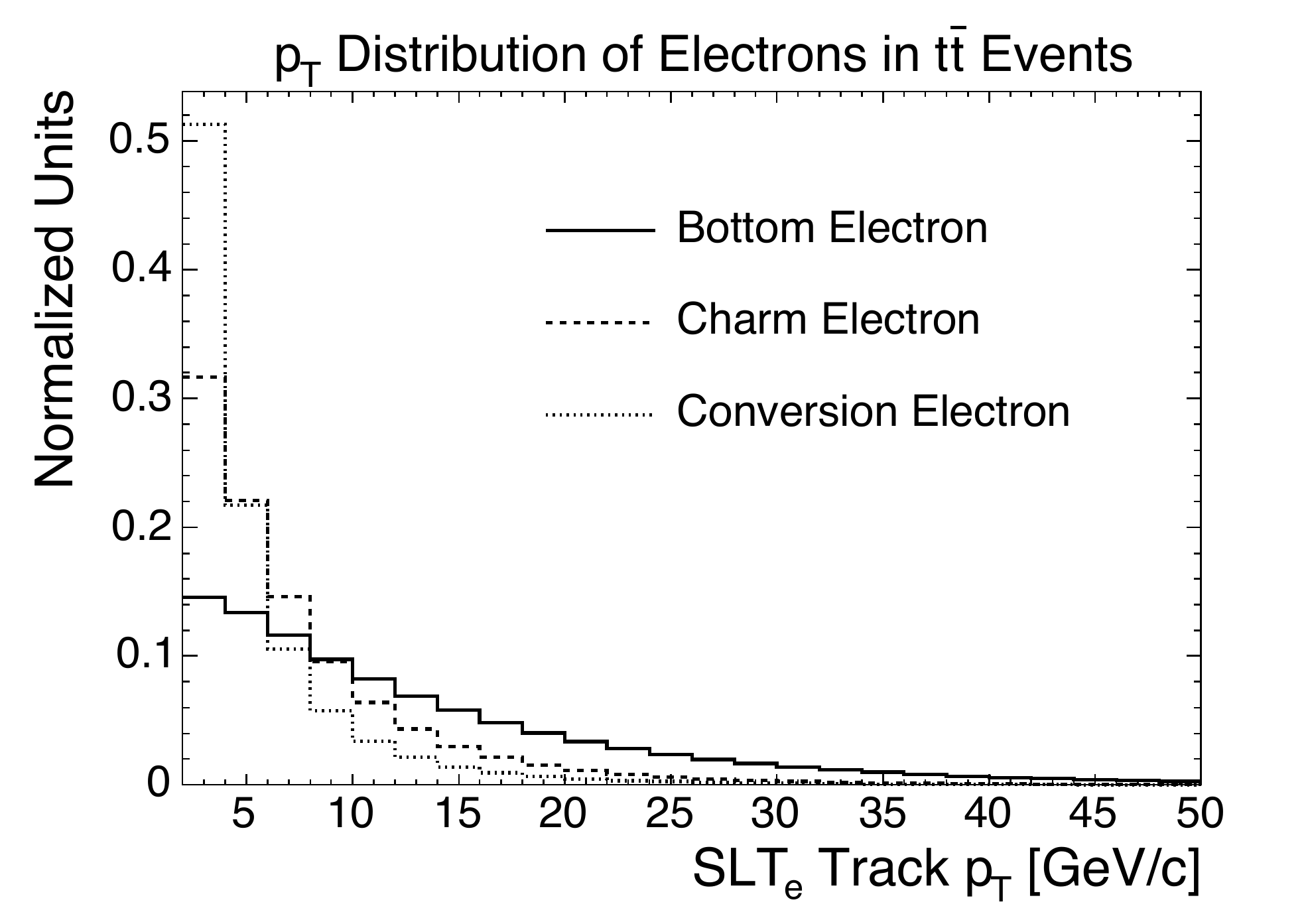}
\caption{Transverse momentum distribution of candidate SLT$_\mathrm{e}$ tracks in jets in \textsc{pythia} $\ttbar$ MC simulated events.  Distributions for electrons from a bottom quark, charm quark, and photon conversions are normalized to unity to emphasize the relative difference in the shapes.} \label{fig:ttbarpt}
\end{center}
\end{figure}

Candidate tracks are passed through the SLT$_\mathrm{e}$ algorithm which uses information from both the calorimeter and CES detectors.  The algorithm is designed to identify low-$\pt$ electrons~\cite{ptfootnote} embedded in high-$\et$ jets while still maintaining a high identification efficiency for high-$\pt$ electrons.  This is particularly important for tagging $\ttbar$ events, although the SLT$_e$ algorithm is not specific to this final state.  Figure~\ref{fig:ttbarpt} shows the $\pt$ shape of candidate SLT$_\mathrm{e}$ electrons in the CDF II detector from a bottom quark, charm quark, and photon conversions in \textsc{pythia} \cite{Pythia} $\ttbar$ Monte Carlo (MC) simulated events.  Even in $\ttbar$ events, the electron spectrum from $b$-jets peaks at low~$\pt$ but extends more than a decade in scale.  Electrons from charm decay in $\ttbar$ events are principally due to cascade decays, but some direct charm production occurs through the hadronic decay of the $W$ boson.

\begin{figure*}[t]
\begin{center}
\includegraphics[width=3.0in]{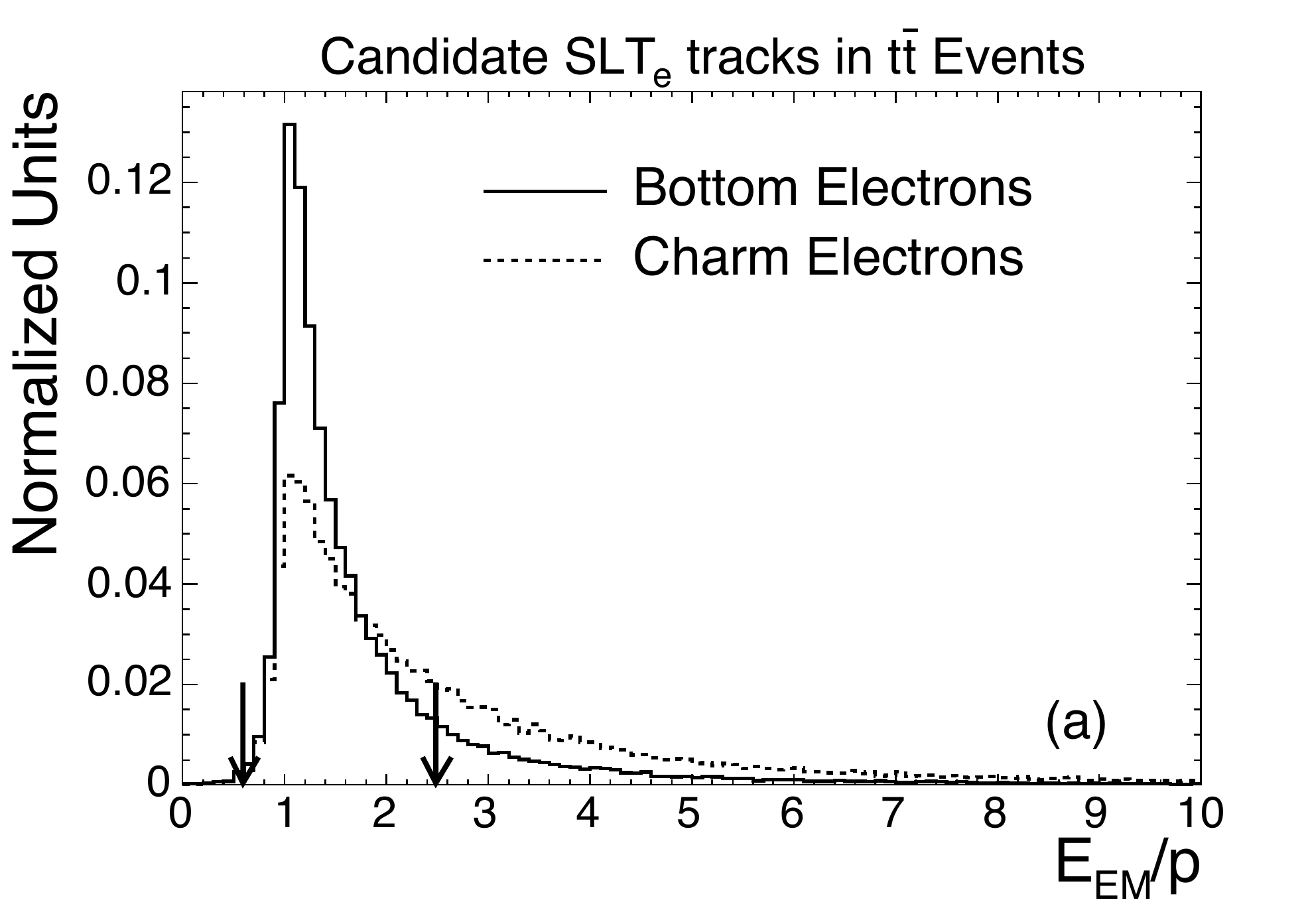}
\includegraphics[width=3.0in]{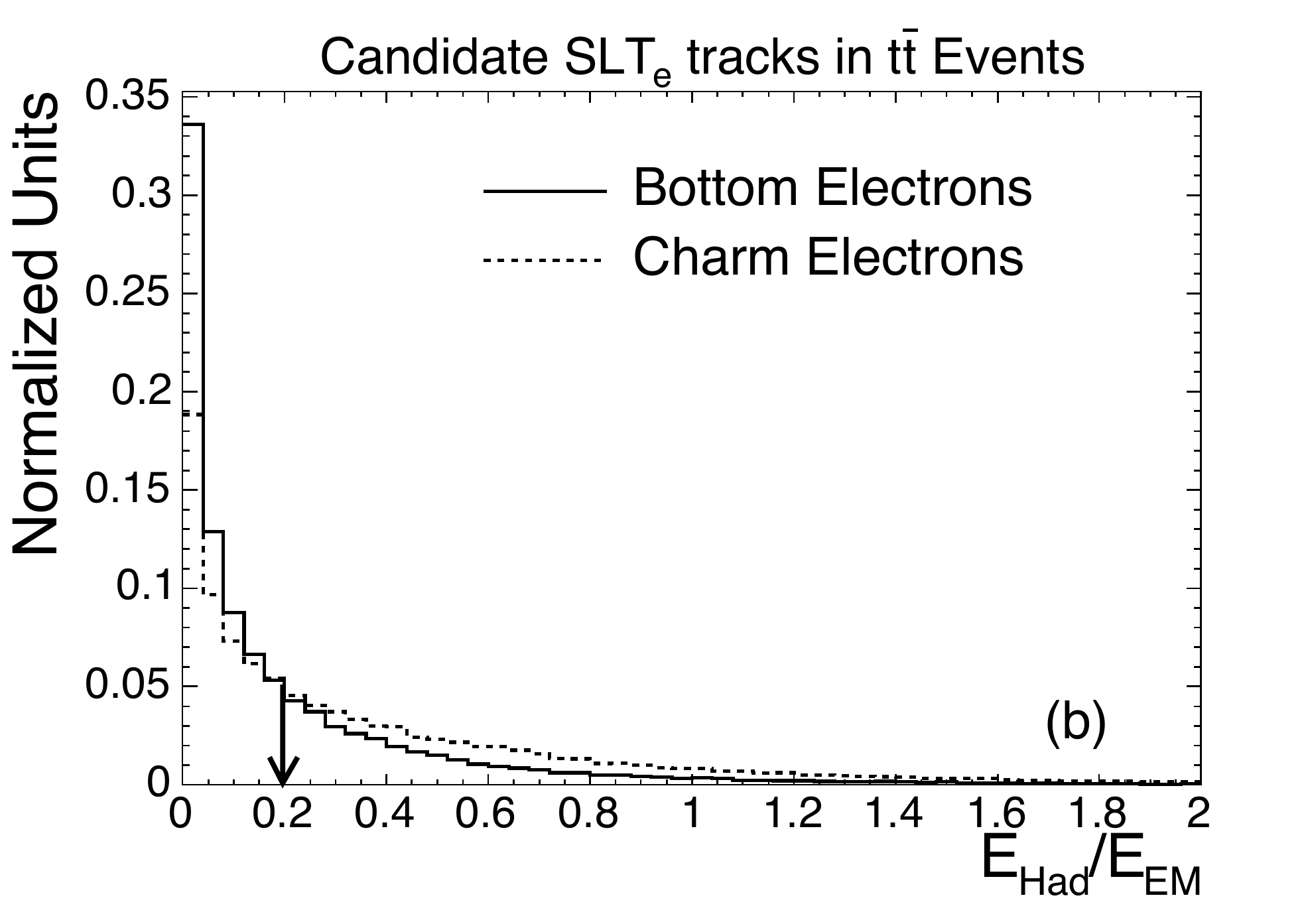}
\caption{(a) $E_{\mathrm{EM}}/p$ and (b) $E_{\mathrm{Had}}/E_{\mathrm{EM}}$ for candidate SLT$_\mathrm{e}$ tracks from HF decay in \textsc{pythia} $\ttbar$ MC simulated events.  Selection criteria for the distributions are shown with arrows.}
\label{fig:cal}
\end{center}
\end{figure*}

The SLT$_\mathrm{e}$ candidate tracks are extrapolated to the front face of the calorimeter to seed an electromagnetic cluster in the CEM.  The two calorimeter towers adjacent in $\eta$-space closest to the extrapolated point are used in the cluster.  A candidate SLT$_\mathrm{e}$ must have an electromagnetic shower that satisfies $0.6<E_{\mathrm{EM}}/p<2.5$ and $E_{\mathrm{Had}}/E_{\mathrm{EM}}<0.2$, where $E_{\mathrm{EM}}$ and $E_{\mathrm{Had}}$ are the total electromagnetic and hadronic energies in the cluster, respectively, and $p$ is the momentum of the electron track.  The $E_{\mathrm{EM}}/p$ requirement selects electromagnetic showers which have approximately the same energy as the track (as expected from electrons), while the $E_{\mathrm{Had}}/E_{\mathrm{EM}}$ requirement suppresses late-developing (typically hadronic) showers.  These requirements were tuned in simulated $\ttbar$ events and are looser than for typical high-$\et$ electrons because the presence of photons and hadrons from the nearby jet distorts the energy deposition.  Figure~\ref{fig:cal} shows the calorimeter variables for candidate SLT$_\mathrm{e}$ tracks in \textsc{pythia} $\ttbar$ simulation.

\begin{figure*}[htb]
\centering
\includegraphics[width=6.0in]{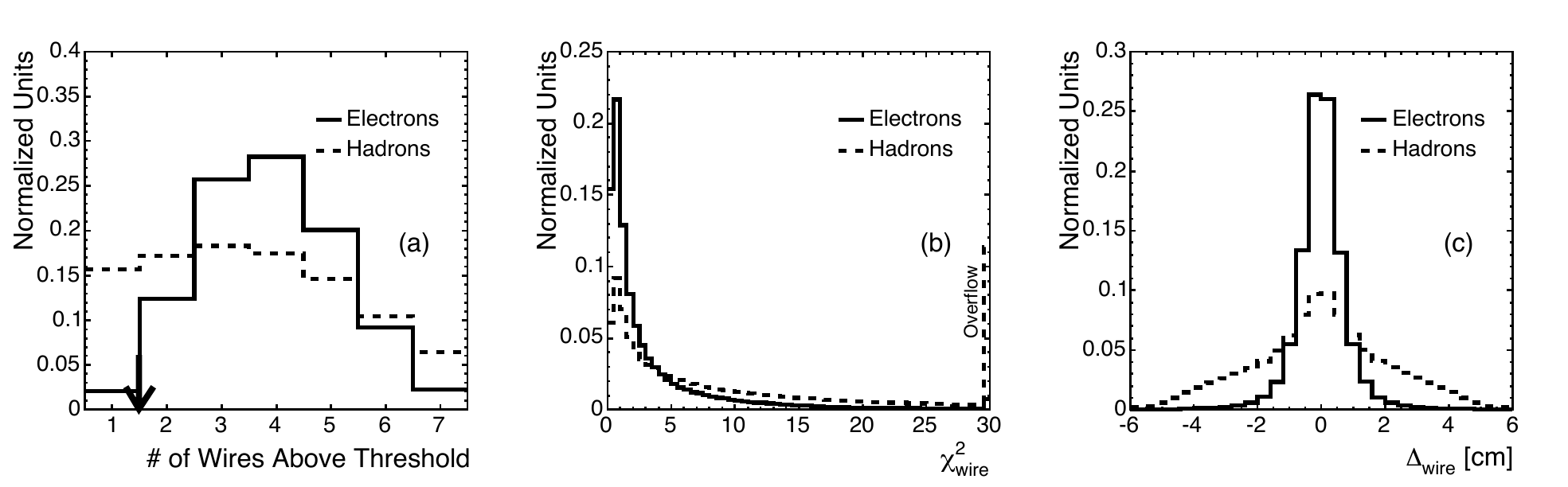}
\includegraphics[width=6.0in]{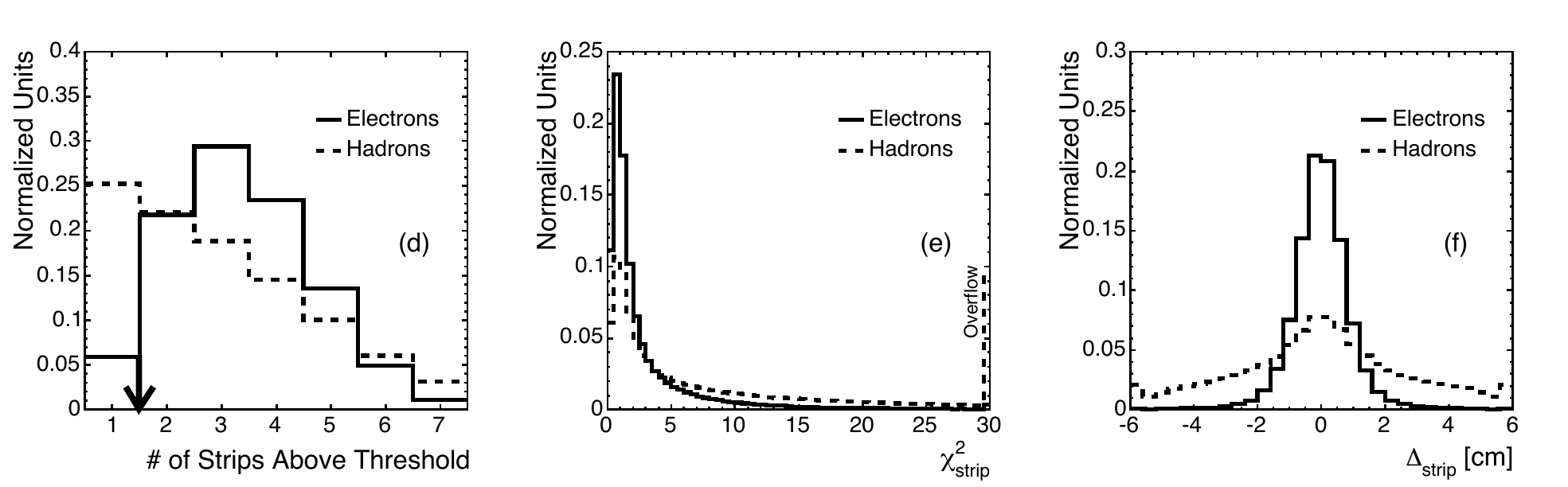}
\caption{(a) Number of wires above threshold, (b) $\chi^2_{\mathrm{wire}}$, (c) $\Delta_{\mathrm{wire}}$, (d) Number of strips above threshold, (e) $\chi^2_{\mathrm{strip}}$, and (f) $\Delta_{\mathrm{strip}}$ for SLT$_\mathrm{e}$ tracks from a sample of conversion electrons and from a sample of hadrons in jet-triggered events.  The last bin of the $\chi^2$ distribution and the first and last bins of the $\Delta$ distribution are the integral of the underflow/overflow.  Arrows indicate the location of the wire and strip requirement for tagging.  CES variables are combined to form a likelihood-ratio discriminant.}
\label{fig:ces}
\end{figure*}

Next, the SLT$_\mathrm{e}$ algorithm uses the track extrapolation to seed a wire cluster and strip cluster in the CES.  We limit the number of strips and wires in the clusters to seven each in order to minimize the effects of the surrounding environment.  At least two wires (strips) with energy above a 80 (120) MeV threshold must be present, or the track is not tagged.  This requirement suppresses low-$\pt$ hadrons that have a late developing shower in the CEM.  Two discriminant quantities determined from the CES are used to distinguish electrons from hadrons.  One is a $\chi^2$ comparison between the transverse shower profile of the SLT$_\mathrm{e}$ candidate and the profile measured with test-beam electrons.  The other is the distance $\Delta$, measured in cm, between the extrapolated track and the position of the cluster energy centroid.  Each type of discriminant is determined for the wire and strip chambers separately.

We construct a likelihood-ratio discriminant by using the $\chi^2$ and $\Delta$ distributions from pure samples of electrons and hadrons as templates.  The electron sample is selected by triggering on a $\et>8$~GeV electron from a photon conversion ($\gamma\to e^+ e^-$) and using the partner electron.  For this sample, the conversion filter requirement is inverted, and the jet-matching requirement is ignored.  To prevent a bias from overlapping electromagnetic showers, photon conversions in which both electrons share a tower are not considered.  The hadron sample is selected through events that pass a 50 GeV jet trigger and identifying generic tracks in jets away from the trigger jet.  In both samples, the purity is over $98\%$.

The distributions for the CES wire chamber and strip chamber discriminants from each sample are shown in Fig.~\ref{fig:ces}.  The relative difference in shapes between the wire and strip distributions is due to the different energy thresholds used, and the slightly different resolution due to the differing technology.

The likelihood-ratio is formed by binning the electron and hadron templates in a normalized 4-dimensional histogram to preserve the correlations between the four variables, $\chi^2_{\mathrm{wire}}$, $\chi^2_{\mathrm{strip}}$, $\Delta_{\mathrm{wire}}$, and $\Delta_{\mathrm{strip}}$, creating probability distribution functions for both signal and background.  We use them to derive a likelihood ratio according to the formula:
\begin{equation}
\mathcal{L}\equiv\frac{S_i}{S_i+B_i}
\end{equation}
where $S_i$ and $B_i$ are the value of the probability distribution functions in the $i^{th}$ bin of signal and background templates, respectively.  We tag a candidate track if $\mathcal{L}>0.55$.  Two other operating points ($>0.65$ and $>0.75$) were also studied for this analysis, but the former point was found to give the best expected combined statistical and systematic uncertainty on the $\ttbar$ cross section.  Table~\ref{table:tag} summarizes the requirements for a candidate SLT$_e$ track to be tagged.

\begin{figure}[htb]
\begin{center}
\includegraphics[width=3.0in]{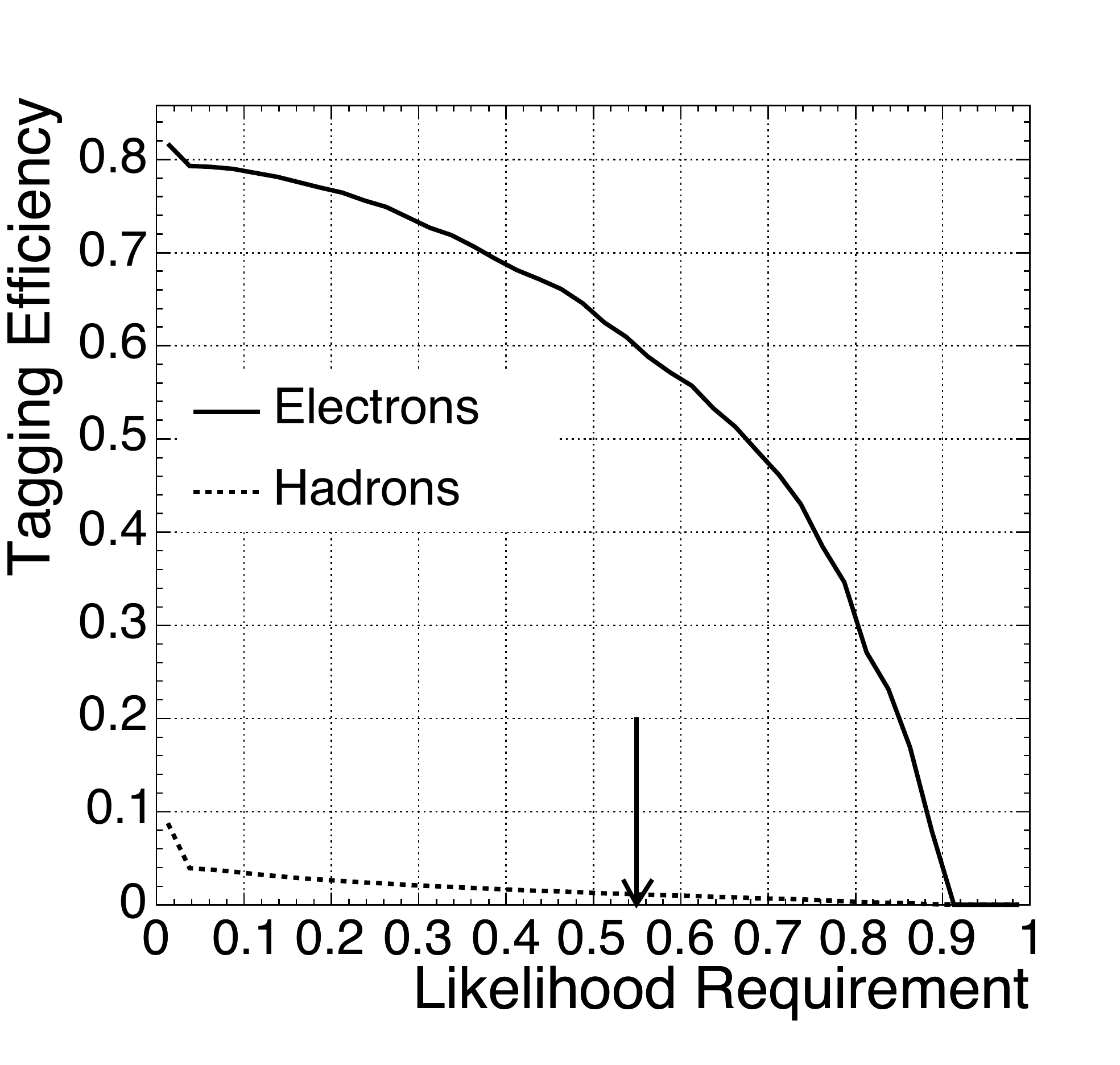}
\caption{Tagging efficiency for electrons from photon conversions (where each leg occupies different calorimeter towers) and hadrons in events triggered on a 50 GeV jet as a function of the likelihood-ratio requirement.}
\label{fig:like}
\end{center}
\end{figure}

We measure the tagging efficiency $-$ that is, the number of tracks that are tagged divided by the number of all candidate tracks $-$ with the combination of calorimeter, wire/strip, and $\mathcal{L}$ requirements in various samples.  Figure~\ref{fig:like} shows this tagging efficiency for the electron sample ($\sim 60\%$ at $\mathcal{L}>0.55$) and the hadron sample ($\sim1.1\%$ at $\mathcal{L}>0.55$) as a function of the likelihood-ratio requirement.  Note that because the hadron sample has not been corrected for the small contamination by electrons, the hadron tagging efficiency should only be considered an upper bound.  This correction is discussed later in Sec.~\ref{sec:hadrons}.  Also note that value of the likelihood-ratio does not extend to 1.0.  This is an artifact of the four variables chosen for the likelihood.  Hadrons occupy the entire phase-space of possible values for $\chi^2_{\mathrm{wire}/\mathrm{strip}}$ and $\Delta_{\mathrm{wire}/\mathrm{strip}}$, so that the background probability distribution function is never zero.

\begin{table}[htb]
\begin{center}
\begin{tabular}{c}
\hline \hline
$0.6<E_{\mathrm{EM}}/p<2.5$ \\
$E_{\mathrm{Had}}/E_{\mathrm{EM}}<0.2$ \\
$\ge2$ wires above threshold in CES cluster \\
$\ge2$ strips above threshold in CES cluster \\
CES $\mathcal{L}>0.55$ \\
\hline \hline
\end{tabular}
\caption{Summary of requirements for tagging a candidate SLT$_e$ track.}
\label{table:tag}
\end{center}
\end{table}

\section{SLT$_\mathrm{e}$ Tagging Efficiency in Jets}\label{sec:efficiency}

An important feature of the SLT$_\mathrm{e}$ algorithm is the tagging efficiency dependence on the environment.  In the previous section we described the per-track tagging efficiency for a sample of isolated conversion electrons where each leg is incident on a different calorimeter tower.  However, the tagging efficiency for electrons from semileptonic $b$ decay with the same kinematic characteristics as the conversion electrons is markedly lower.  This is due to the nearby jet which distorts the electromagnetic shower detected in the calorimeter.  In general, the calorimeter variables, $E_{\mathrm{EM}}/p$ and $E_{\mathrm{Had}}/E_{\mathrm{EM}}$, are strongly affected by the jet, whereas the CES variables - that is the $\chi^2$ and $\Delta$ variables as well as the number of wires and strips in the CES cluster - have a much weaker dependence.

For the SLT$_\mathrm{e}$ algorithm, we introduce the isolation variable, $I_\mathrm{SLT}$, defined as the scalar sum of the $\pt$ of tracks which point to the calorimeter cluster divided by the candidate track $\pt$: $\Sigma_{\mathrm{clst}}\pt/\pt$.  This variable is useful at quantifying the degree to which the local environment should affect the electron's electromagnetic shower, and hence the identification variables.  An isolated SLT$_\mathrm{e}$ track has $I_\mathrm{SLT}$ identically equal to 1.0, whereas for a non-isolated track, $I_\mathrm{SLT}>1.0$.

In order to measure the SLT$_\mathrm{e}$ tagging efficiency of soft electrons in jets, we rely on a combination of MC simulation and data-driven techniques.  We study the calorimeter and the CES discriminants, which both enter the SLT$_\mathrm{e}$ algorithm, separately.  Although the calorimeter variables have a strong dependence on the local environment, they are well-modeled in the MC simulation.  However, the CES variables, on the whole, are poorly modeled in the simulation due to the presence of early overlapping hadronic showers.

\begin{figure*}[htb]
\begin{center}
\includegraphics[width=3.0in]{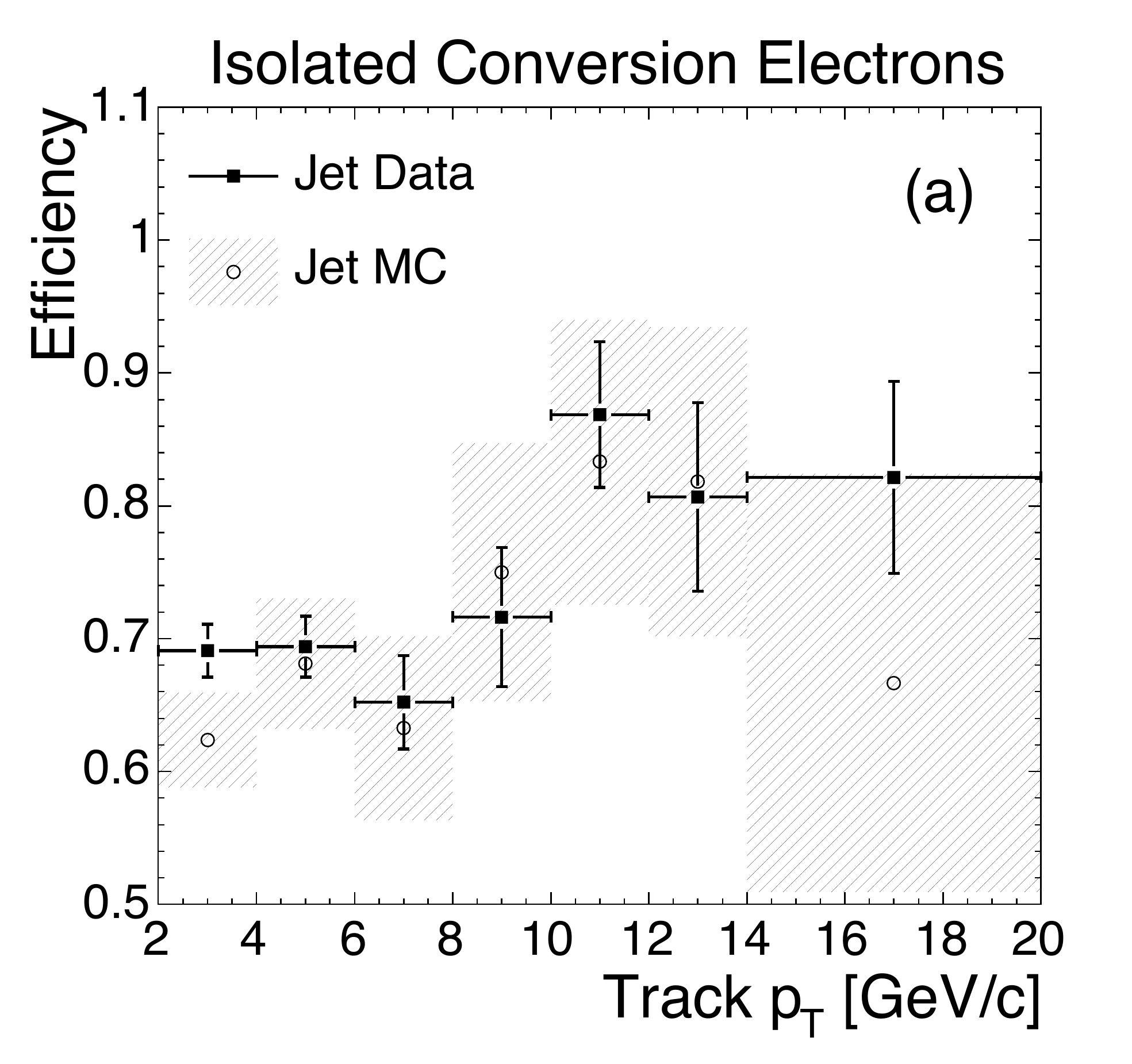}
\includegraphics[width=3.0in]{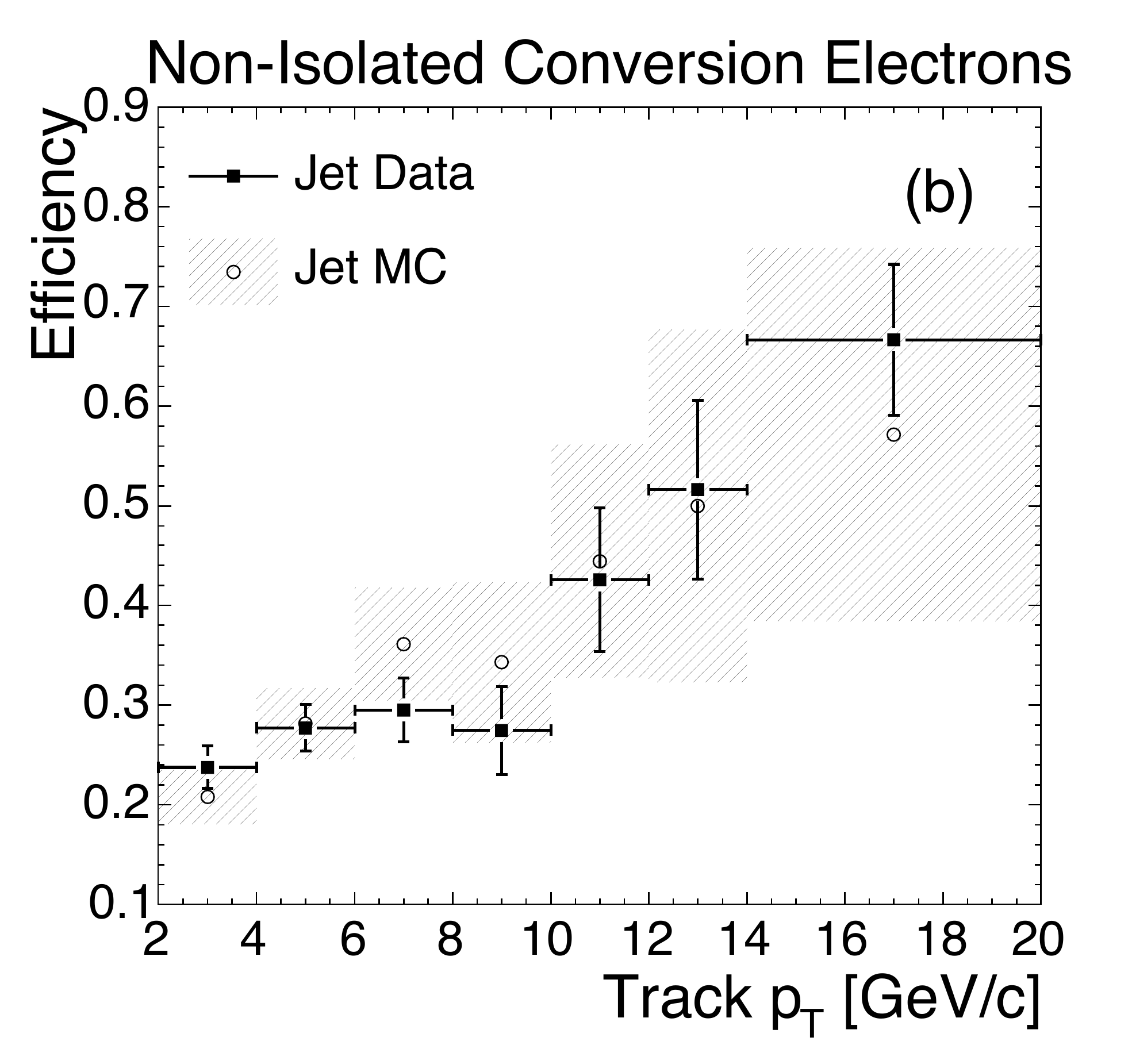}
\caption{Efficiency of the calorimeter requirements on an untagged conversion electron as a function of the track $\pt$, for both isolated (a) and non-isolated (b) tracks.  Error bars reflect statistical uncertainties from both data and MC.  We use the overall agreement to derive a 2.5\% relative systematic uncertainty on the calorimeter requirements of the SLT$_\mathrm{e}$ tagger.}
\label{fig:isolconv}
\end{center}
\end{figure*}

We study the modeling of the SLT$_\mathrm{e}$ calorimeter-based discriminants in a sample of conversion electrons reconstructed in jets.  This sample is constructed by identifying an electron and its conversion partner while both are close to a jet ($\Delta R\le 0.4$).  We select such conversions in data triggered on a 50 GeV jet and a kinematically comparable dijet MC simulation sample.  We use the missing silicon layer variable, described in Sec.~\ref{sec:conversions}, to enhance the conversion electron content in the sample.  This is done by requiring that the track associated with the conversion partner is expected to have, but does not have, hits in at least three silicon layers.  The conversion partner is used as a probe to compare the efficiency of the combined calorimeter requirements in both data and simulated samples as a function of $\pt$ and $I_\mathrm{SLT}$.  We see very good agreement in the general trend between both samples, as shown in Fig.~\ref{fig:isolconv}, from which we derive a 2.5\% relative systematic uncertainty (integrating over all bins) to cover the difference between data and simulation.  The comparison between kinematically and environmentally similar samples is important to validate the behavior of the simulation modeling.

To account for the mis-modeling of the CES-based discriminants, we measure the tagging efficiency of candidate SLT$_\mathrm{e}$ tracks directly in data and apply it to candidate SLT$_\mathrm{e}$ tracks in the simulation that have already passed the calorimeter-based requirements.  The efficiency is parameterized as a three-dimensional matrix in $\pt$, $\eta$, and $I_\mathrm{SLT}$ to account for the correlations between the three variables.  This matrix is constructed out of the pure conversion electron sample used to create the likelihood-ratio templates.  The validity of the tag matrix is then verified in a sample of electrons from $Z$ boson decays and in a $\bbbar$ sample, as described in Sec.~\ref{sec:bbbar}.  A 3\% relative systematic uncertainty $-$ derived from the agreement within the conversion sample and with the $Z\rightarrow e^+ e^-$ sample $-$ is applied to the tag-matrix prediction.

Applying the matrix as a weight on each candidate SLT$_\mathrm{e}$ track identified in the simulated events, we find that the tagging efficiency for electrons from HF jets in $\ttbar$ events is approximately 40\% per electron track (See Sec.~\ref{sec:bbbar}).  This is calculated by identifying candidate SLT$_\mathrm{e}$ tracks in $\ttbar$ events matched to electrons from HF jets in the simulation.  For those electrons which pass the calorimeter requirements, the tag matrix determines the expected tagging probability.

\section{SLT$_\mathrm{e}$ Tagging Backgrounds}\label{sec:fakes}

The two principal backgrounds to SLT$_\mathrm{e}$ tagging are real electrons from photon conversions and misidentified electrons from charged hadrons (\textit{e.g.} $\pi$, $K$, $p$).  Although the tagging probability is very low for hadrons, the high multiplicity of such tracks makes their contribution non-negligible.  Conversion electrons are much more abundant than electrons from HF jets.  In $\ttbar$ events, three times as many candidate tracks are due to conversion electrons than to electrons from semileptonic decay of HF.  Their removal is essential to effective $b$-tagging.  Additionally, there is a small contribution from Dalitz decays of $\pi^0$, $\eta$, and $J$/$\psi$.  In this section we discuss the estimation of the conversion electron and hadronic backgrounds.

\subsection{Conversions}\label{sec:conversions}

The primary procedure for conversion electron rejection relies on identifying the partner leg.  We identify an SLT$_\mathrm{e}$ tagged track as a conversion if, when combined with another nearby track in the event, the pair has the geometric characteristics of a photon conversion.  In particular, the $\Delta\cot(\theta)$ between the tracks as well as the distance between the tracks when they are parallel in the $r-\phi$ plane must be small.  However, for low-$\pt$ conversion electrons in jets, this requirement fails to identify the partner leg more than 40\% of the time.  The primary reason for this is that the track reconstruction algorithms begin to fail at very low $\pt\sim500$~MeV/$c$.  The asymmetric energy sharing between conversion legs exacerbates this effect.

\begin{figure}[t]
\begin{center}
\includegraphics[width=3.0in]{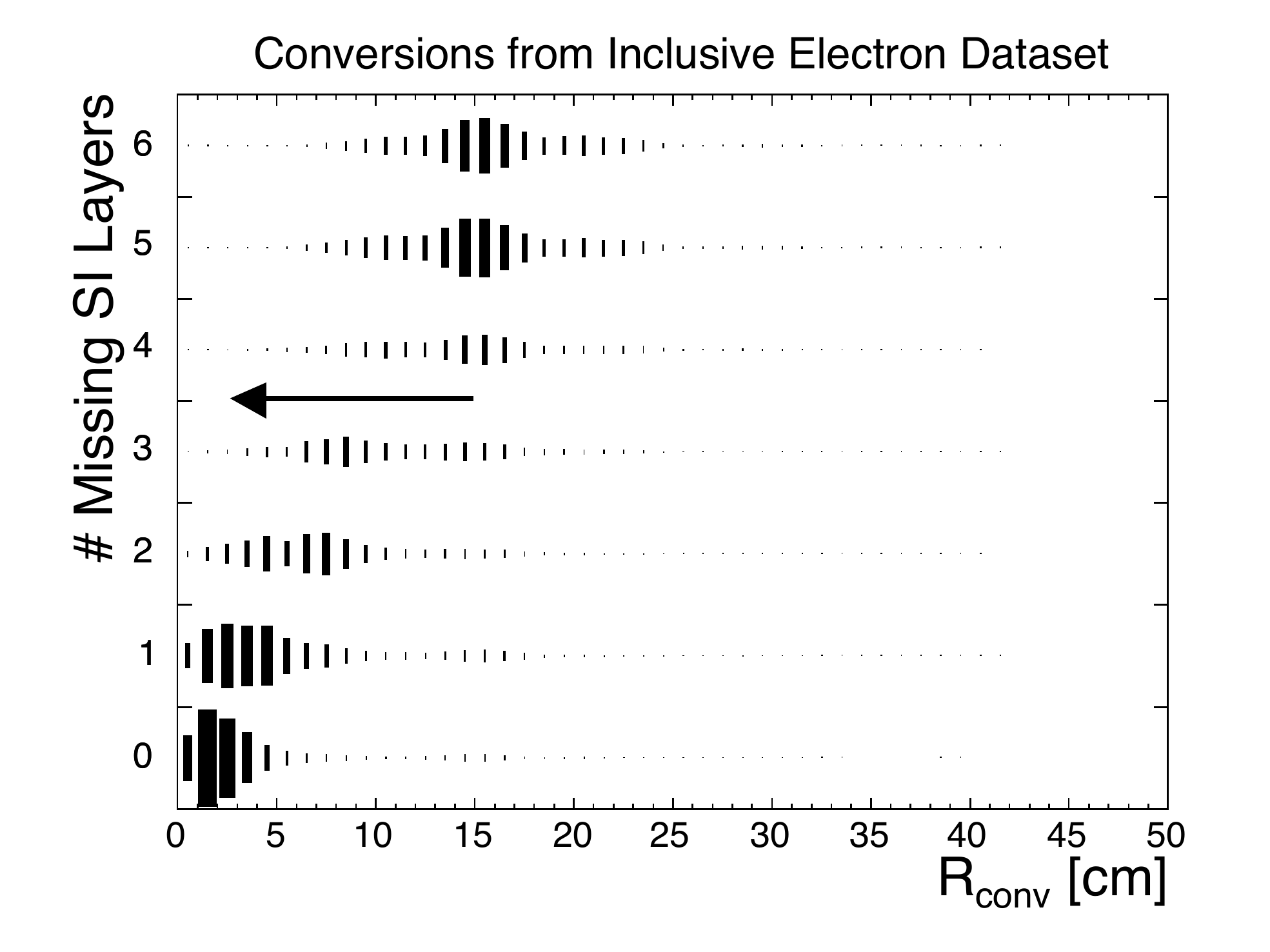}
\caption{Number of missing silicon layers versus the reconstructed radius of conversion for conversion electrons found in an inclusive ($\et>8$~GeV) electron sample.  Tracks tagged by the SLT$_\mathrm{e}$ algorithm are rejected as conversions if they have more than three missing silicon layers.}
\label{fig:conv}
\end{center}
\end{figure}

To recover conversion electrons when the partner leg is not found, we use the fact that conversions are produced through interactions in the material.  We extrapolate the candidate track's helix through the silicon detectors and identify silicon detector channels where no hit is found.  If a track is missing hits on each side of more than three double-sided silicon layers, then it is identified as a conversion (at most six missing layers is possible~\cite{L00Removal}).  Figure~\ref{fig:conv} shows the reconstructed radius of conversion, $R_\mathrm{conv}$, versus the number of missing silicon layers for conversion electrons with both legs tagged by the SLT$_\mathrm{e}$ in an inclusive sample of $\et>8$~GeV electrons.  Although high $R_\mathrm{conv}$ values are suppressed because of the impact parameter requirement, there is a clear correlation between missing silicon layers and the $R_\mathrm{conv}$.  For SLT$_\mathrm{e}$ candidates, we combine both the standard partner-track-finding algorithm with the missing silicon layer algorithm so that if a tag fails either, we reject it as a conversion electron.

We measure the conversion ID efficiency in data by decomposing the algorithm into a partner-track-finding component and a missing silicon layer component.  We use the missing silicon layer templates to measure the partner-track-finding component efficiency, and we use a sample of conversions with both legs SLT$_\mathrm{e}$ tagged to measure the missing silicon layer component efficiency.  We combine the efficiencies, accounting for their correlation.

We use an \textit{in situ} process of building templates for the missing silicon layer variable for conversions and prompt tracks directly within the sample of interest to fit for the total conversion content before and after rejection.  The \textit{in situ} nature of the template construction is important because conversion identification depends strongly on kinematics and geometry that can vary across different samples.  The conversion template is constructed from conversions where both legs are tagged by the SLT$_\mathrm{e}$, and the prompt track template is constructed from tracks where the SLT$_\mathrm{e}$ requirements have been inverted, resulting in a nearly 100\% pure hadronic sample.

\begin{figure}[t]
\begin{center}
\includegraphics[width=3.0in]{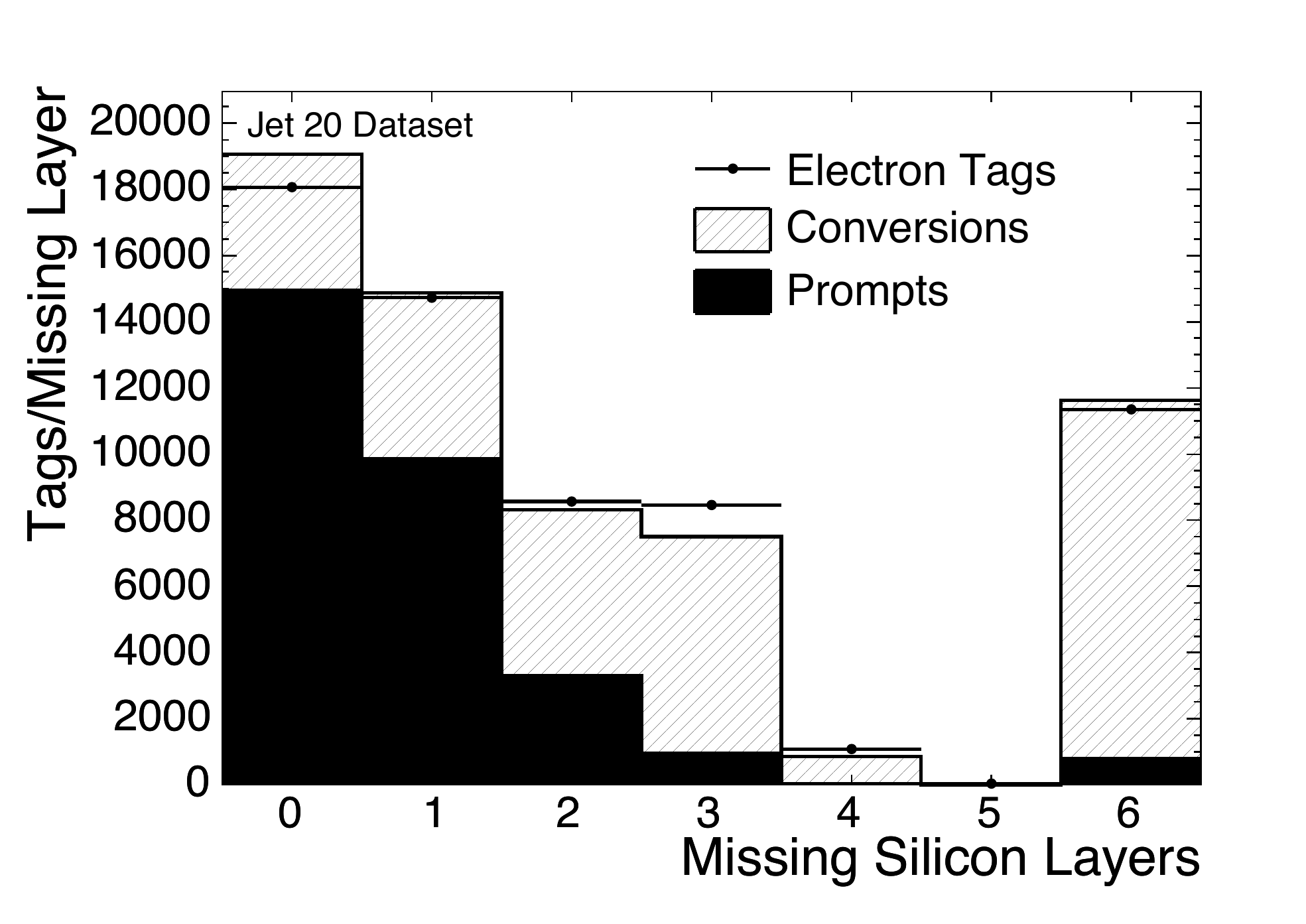}
\caption{Fit for the conversion and prompt component of SLT$_\mathrm{e}$ tags before conversion removal in events triggered on a $\et>20$~GeV jet.  The goodness-of-fit is limited by systematic biases in the template construction and is accounted for in the final SF measurement.}
\label{fig:convfit}
\end{center}
\end{figure}

A fit for the conversion component of SLT$_\mathrm{e}$ tags in events triggered on a 20~GeV jet is shown in Fig.~\ref{fig:convfit}.  In the fit, only those electrons with hits in six expected silicon layers is considered.  Those tracks with fewer than six expected layers are used as a consistency check, and a systematic uncertainty is assigned to the geometric bias incurred from this requirement.  The dearth of tracks with four or five missing layers is an artifact of the CDF track reconstruction algorithm, which requires that at least three silicon hits must be added to any track or none will be added.  The goodness-of-fit is limited by systematic biases in the template construction which contribute the dominant systematic uncertainties to the efficiency measurement.  Such biases include correlations between track finding and missing silicon layers, modeling of prompt electrons (including HF decay) by prompt hadrons, geometric dependencies, and sample contamination.

\begin{figure}[t]
\begin{center}
\includegraphics[width=3.0in]{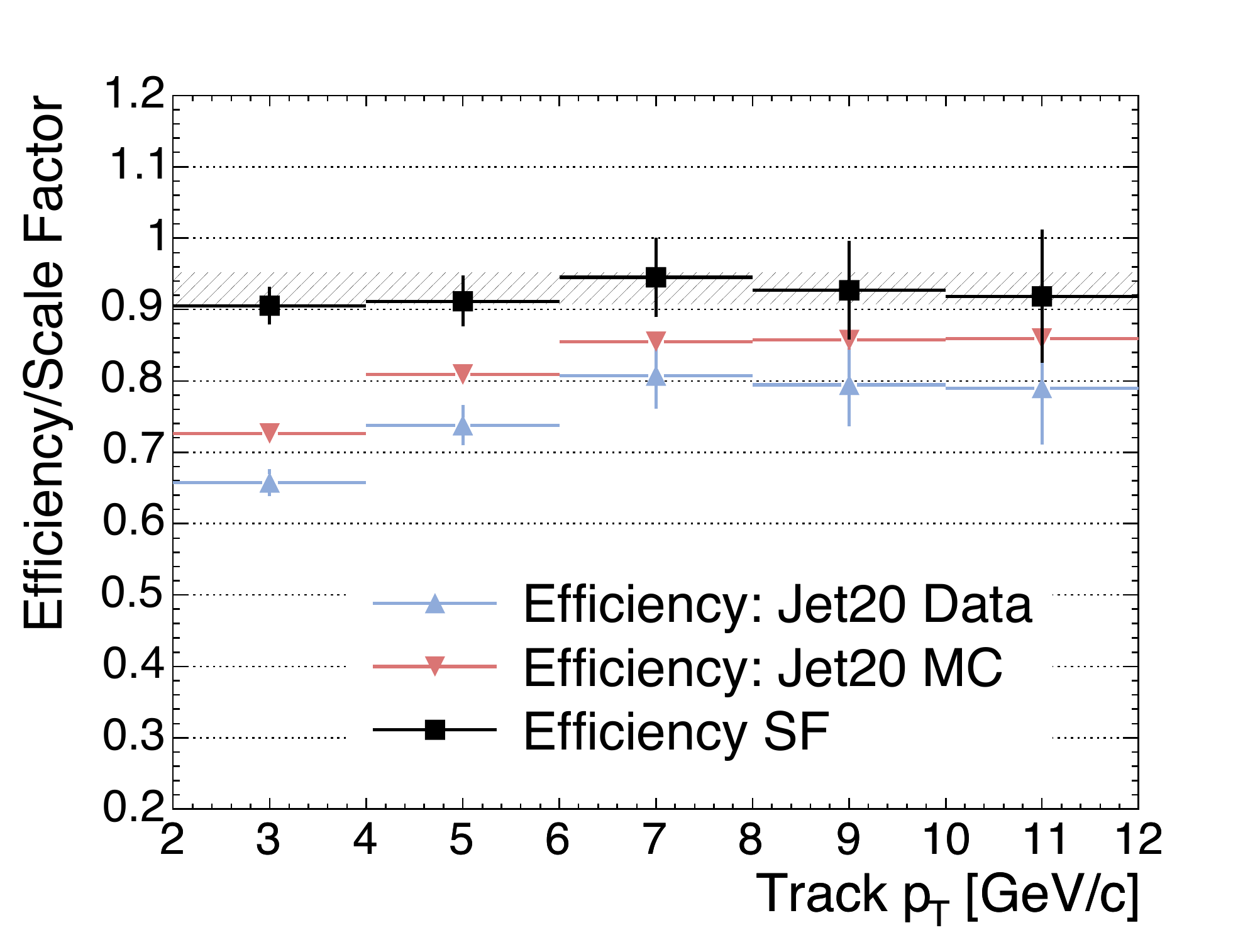}
\caption{Conversion identification scale factor measurement in events triggered by a $\et>20$~GeV jet.  Shown also is the efficiency measured in data and the efficiency measured in MC simulation.  The consistency across $\pt$, among other variables, demonstrates the validity of the SF approach.}
\label{fig:convSF}
\end{center}
\end{figure}

We find that the conversion identification efficiency is overestimated in MC simulation relative to data.  We characterize the difference by a multiplicative scale factor (SF), defined as the ratio of efficiencies measured in data and simulation.  Because the conversion identification efficiency depends strongly on the underlying photon energy spectrum, it is important for the SF measurement to compare energetically similar samples.  Therefore, we measure the SF in events triggered by a jet with $\et>20$, 50, 70, and 100 GeV and compare to MC simulated dijet events which pass the same requirements.  We measure a conversion efficiency SF of $0.93\pm0.01$\,(stat)\,$\pm0.02$\,(syst).  The dominant uncertainties are systematic effects related to the accuracy of the template models.  We find that the SF behaves consistently as a constant correction across a variety of different event and track variables in multiple datasets.  Figure~\ref{fig:convSF} shows the SF as a function of track $\pt$ in a sample of events triggered by a $\et>20$~GeV jet.  The gray band shows the value of the SF with statistical and systematic uncertainties for combined SF across jet~20, 50, 70, and 100 datasets.

We also measure a conversion `misidentification efficiency' $-$ defined as the efficiency to misidentify a non-conversion track as a conversion $-$ multiplicative SF of $1.0\pm0.3$ between data and simulation.  This is done by measuring the efficiency to identify prompt tracks as conversions.  The large systematic uncertainty accounts for the variation found across different kinematic variables, jet triggers, and particle types (such as the difference between a HF electron and a pion from $K_s$ decay).  In $\ttbar$ events, the complete algorithm is approximately 70\% efficient at rejecting candidate SLT$_\mathrm{e}$ tracks that are conversions.  Only 7\% of non-conversions are misidentified as conversions.  Since the misidentification efficiency is an order of magnitude lower than the efficiency, the total contribution of systematic uncertainties from each is comparable.

\subsection{Hadrons}\label{sec:hadrons}

\begin{figure}[t]
\begin{center}
\includegraphics[width=3.0in]{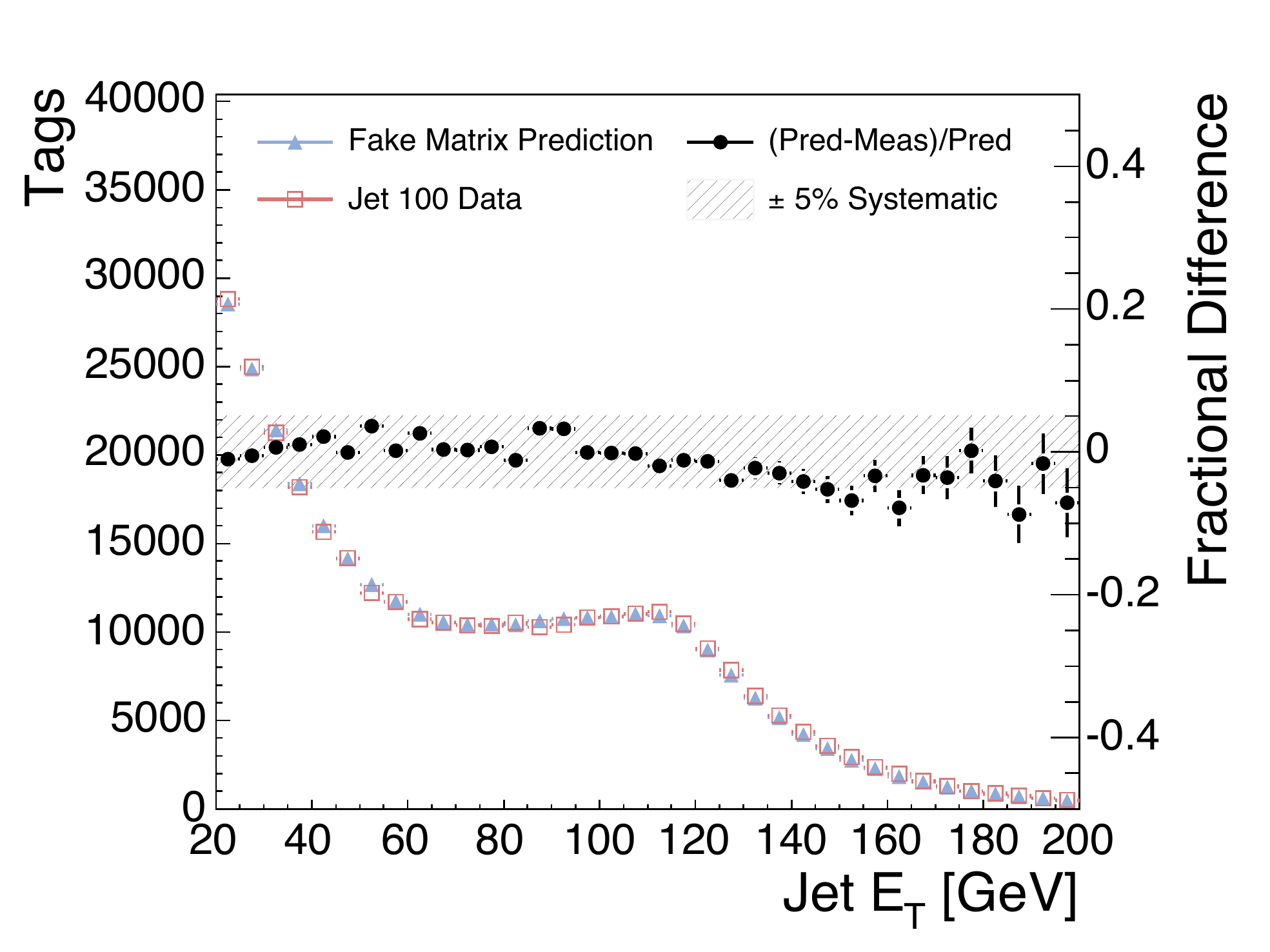}
\caption{Predicted and measured tags in events triggered on a $\et>100$~GeV jet as a function of the $\et$ of the jet closest to the candidate SLT$_\mathrm{e}$ track.  On the right axis is the relative fractional difference between the measurement and prediction.}
\label{fig:fake}
\end{center}
\end{figure}

We measure the tagging efficiency of hadrons in MC by defining a three-dimensional fake matrix out of tracks in jet-triggered events.  The matrix parameterizes the probability that the CES discriminants ($\mathcal{L}$ plus the number of wires and strips) can tag a hadron.  We remove jets where a large fraction of energy is deposited by a single track, in order to reduce the contamination of hard electrons that are also reconstructed as jets.  We find that the use of the track $\pt$, $\eta$, and $I_{\mathrm SLT}$ are sufficient to describe the dependence of the tagging efficiency on other variables as well.  This is demonstrated in Fig.~\ref{fig:fake}, which shows the measured and predicted tags in events triggered on a jet with $\et>100$~GeV as a function of the $\et$ of the jet closest in $\Delta R$ to the SLT$_\mathrm{e}$ track.  We also cross-check the fake-matrix prediction in a distinct sample of tracks in jets triggered on a high-$\et$ photon.  We find that the agreement is good within $\pm 5\%$.

Before tagging, the tracks in the jet samples are almost purely hadronic; however, after tagging, we must correct for the electron contamination when we estimate the total efficiency to tag a hadronic track.  Three classes of electrons are present in the sample: conversion electrons, HF electrons, and other sources (primarily Dalitz decay of $\pi^0$).  The conversion electron contamination is estimated by measuring the efficiency and misidentification efficiency of the conversion filter in the jet samples.  Using this information in combination with the number of tracks before and after conversion removal determines the remaining conversion content.  The HF electron contamination is estimated using correlations between the SLT$_\mathrm{e}$ tags and $b$-tags from a secondary vertex algorithm, \textsc{SecVtx}~\cite{Secvtx}.  We use \textsc{SecVtx} to enhance the HF content of the jet sample.  Using MC simulation to estimate the expected size of this enhancement, we can extrapolate back to the original, pre-\textsc{SecVtx} tag HF component.  The remaining contribution of electrons from other sources is small and estimated with the MC simulation.

We find that $35\pm3\%$ of the tags in the jet-triggered sample are electrons.  This estimate is verified by measuring the SLT$_\mathrm{e}$ tagging efficiency for charged pions from $K_s$ decay.  By subtracting the electron contamination from the fake-matrix prediction, we find that, on average, 0.5\% of hadronic tracks in $\ttbar$ events produce a fake SLT$_\mathrm{e}$ tag.

\section{Tuning in $\bbbar$ Sample}\label{sec:bbbar}

As a validation of the measured efficiency of the tagger, we measure the jet tagging efficiency in a highly enriched sample of $\bbbar$ events.  Events are selected through an 8~GeV electron or muon trigger, and we require that both the jet close to the lepton ($\Delta R\le 0.4$) and the recoiling (away) jet have a \textsc{SecVtx} tag.  We measure the per-jet efficiency to find at least one SLT$_\mathrm{e}$ tag in the away jet.  This efficiency is measured to be $4.4\pm0.1$\,(stat) (\%) in simulation and $4.3\pm0.1$\,(stat) (\%) in data.

\begin{figure}[h]
\begin{center}
\includegraphics[width=3.0in]{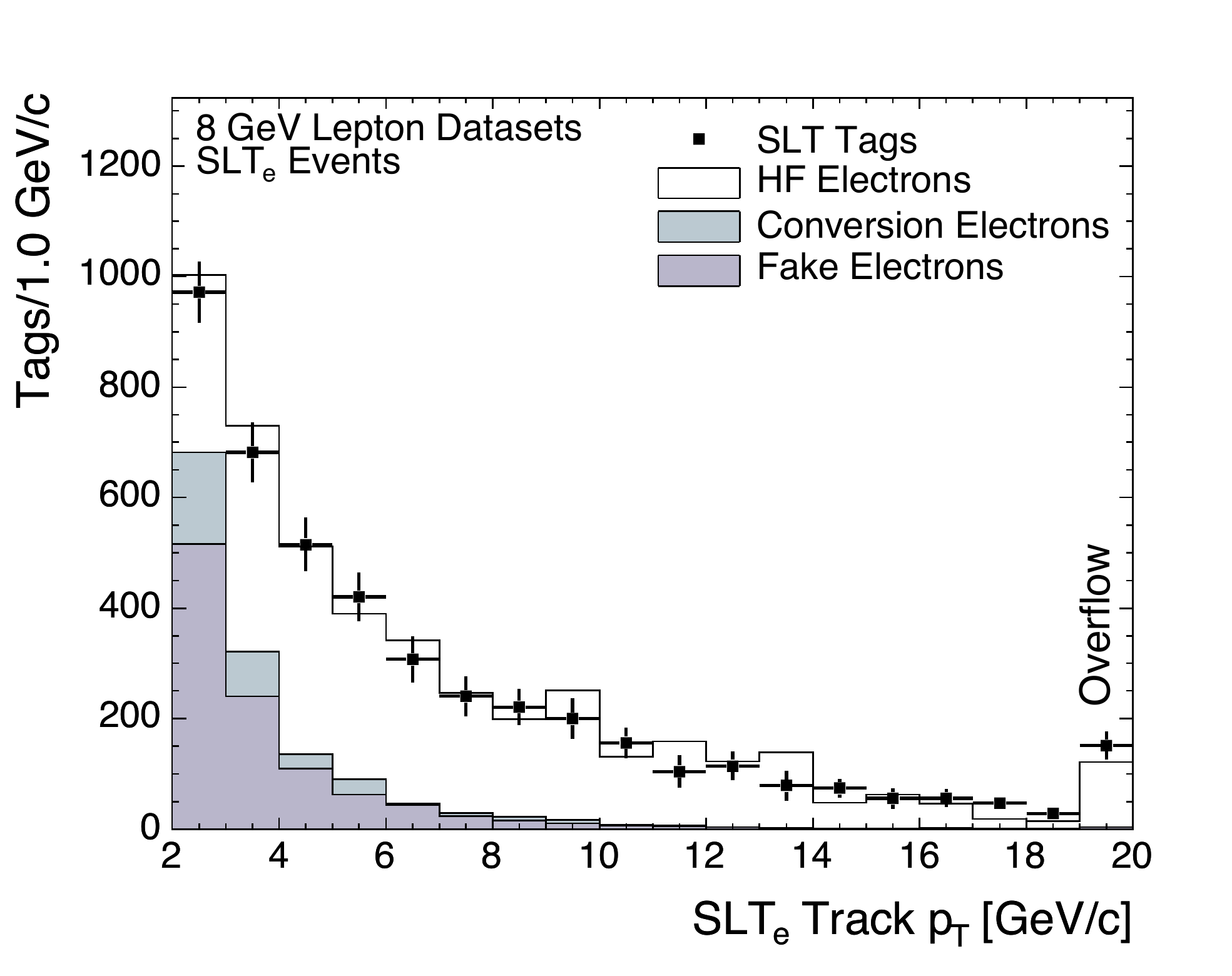}
\caption{Predicted and measured tags as a function of the SLT$_\mathrm{e}$ track $\pt$ in a $\bbbar$ enhanced sample constructed from inclusive electron and muon triggered events.  Shown are contributions from fake tags, conversion electron tags, and HF electron tags. Simulation and data statistical uncertainties are combined in quadrature and shown together on the data points only.}
\label{fig:bbbar}
\end{center}
\end{figure}

\begin{figure*}[!t]
\begin{center}
\includegraphics[width=3.0in]{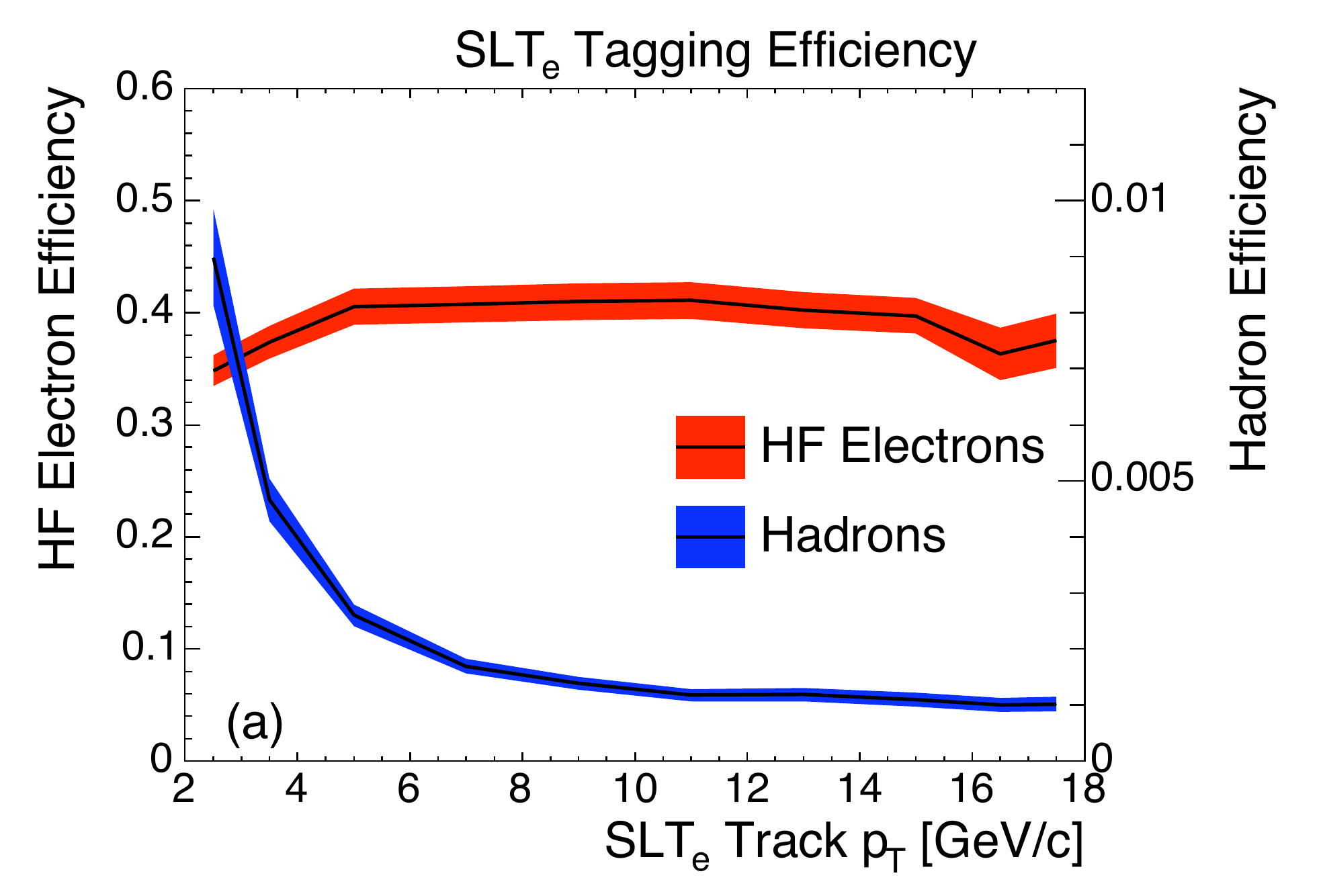}
\includegraphics[width=3.0in]{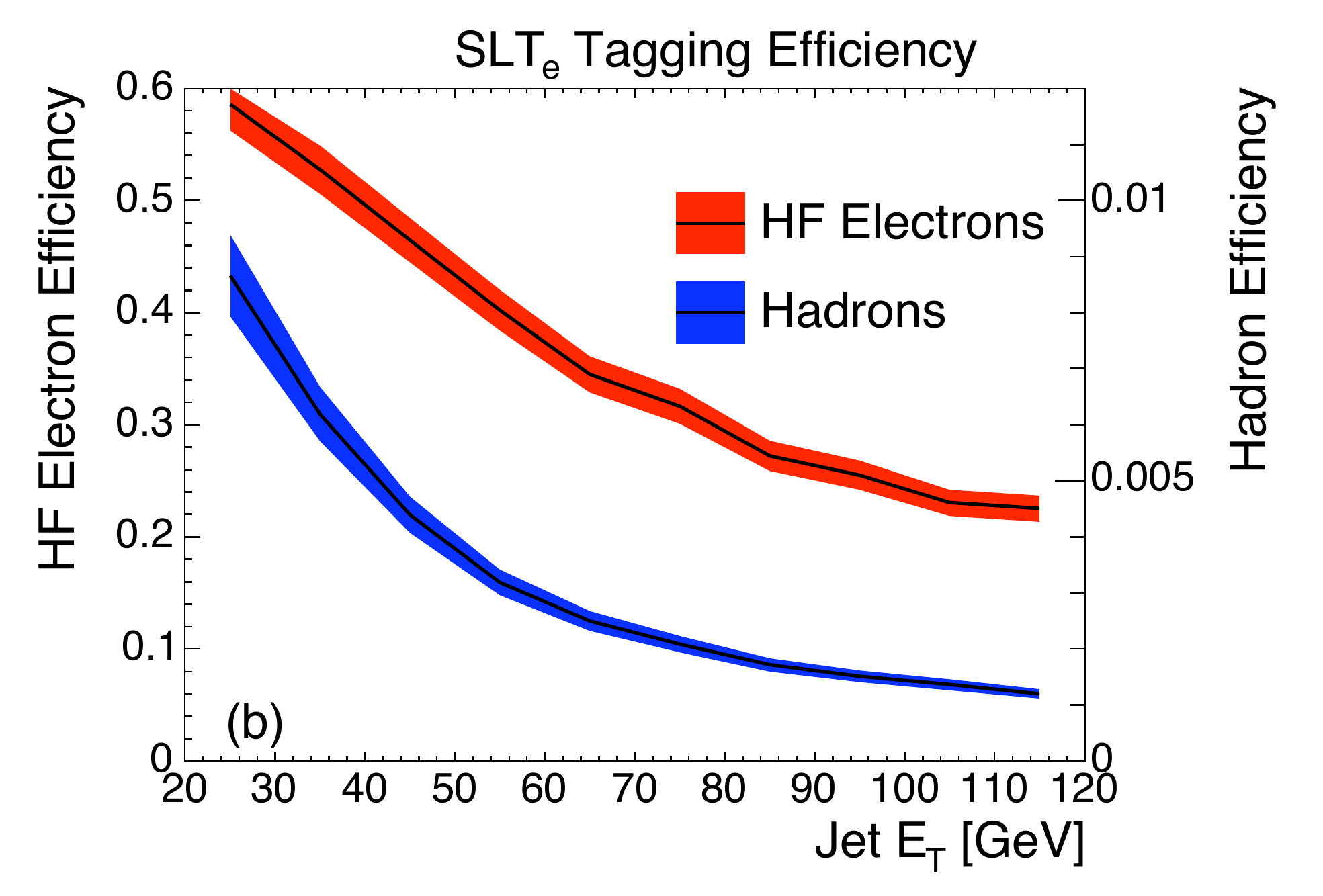}
\caption{Predicted efficiency to tag an electron from semileptonic decay of HF and a hadron candidate SLT$_\mathrm{e}$ track in $\ttbar$ events as a function of the track $\pt$ (a) and corrected jet $\et$ (b).  The left axis indicates the tagging efficiency for the electrons and the right axis indicates the tagging efficiency for the hadrons.}
\label{fig:MCeff}
\end{center}
\end{figure*}

\begin{table*}[!t]
\begin{center}
\begin{tabular}{cccccc}
\hline
\hline
\multicolumn{6}{c}{Corrected $\ttbar$ Acceptance (\%)} \\
\hline
Lepton & 1 jet & 2 jets & 3 jets & 4 jets & $\ge 5$~jets \\
\hline
CEM & $0.163\pm0.003$ & $0.862\pm0.011$ & $1.403\pm0.017$ & $1.493\pm0.018$ & $0.519\pm0.007$ \\
CMUP & $0.089\pm0.002$ & $0.477\pm0.009$ & $0.788\pm0.015$ & $0.826\pm0.015$ & $0.284\pm0.006$ \\
CMX & $0.042\pm0.001$ & $0.220\pm0.005$ & $0.353\pm0.008$ & $0.381\pm0.008$ & $0.130\pm0.003$ \\
\hline
Total & $0.295\pm0.005$ & $1.559\pm0.024$ & $2.543\pm0.039$ & $2.700\pm0.041$ & $0.932\pm0.015$ \\
\hline
\hline
\end{tabular}
\caption{Corrected $\ttbar$ acceptance in the lepton+jets decay channel.  We have required $\Ht>250$~GeV for events with $\ge3$~jets and $\displaystyle{\not}{\et}>30$~GeV.  Combined statistical and systematic uncertainties are shown.}
\label{table:acc}
\end{center}
\end{table*}

The efficiency is calculated in simulation by taking all of the candidate tracks in the jet that pass the calorimeter requirements and using either the tag matrix for electrons or the fake matrix for hadrons to determine a tagging probability.  If a track is identified as a conversion, then the tagging probability is rescaled according to the conversion efficiency or mis-identification efficiency SF.  We tune the tag matrix with a multiplicative factor of $0.98\pm0.03$ to get the simulation to agree with data, where the systematic uncertainty is assigned to cover a jet-$\et$ dependence in the difference.  The difference in the prediction and measurement is due to isolation effects of the jet environment not already accounted for by the $I_\mathrm{SLT}$ parameterization, specifically the presence of neutral hadrons.  Figure~\ref{fig:bbbar} shows the predicted and measured tags in the combined 8~GeV electron and muon trigger samples as a function of the $\pt$ of the SLT$_\mathrm{e}$ tag after the tuning.  Statistical uncertainties from the data and simulation are added in quadrature and shown on the data points.

By combining the tag matrix, fake matrix, conversion identification and mis-identification efficiency SFs, and the correction for the jet environment, we estimate the tagging efficiency of data from simulation.  Figure~\ref{fig:MCeff} shows the efficiency to tag a HF electron and a hadron in simulated $\ttbar$ events as a function of the track $\pt$ and the jet $\et$.    While the tagging efficiency for electrons is steady as a function of the track $\pt$, it decreases as a function of the jet $\et$ because of the decreasing isolation at high-$\et$.

\section{Cross Section Measurement}\label{sec:xs}

The $\ttbar$ production cross section is determined with the equation
\begin{equation}
\sigma=\frac{N-B}{\epsilon_{\ttbar}\,\mathcal{A}_{\ttbar}\int\mathcal{L}dt}\label{eq:xs}
\end{equation}
where $N$ is the number of tagged events, $B$ is the expected background, $\epsilon_{\ttbar}$ and $\mathcal{A}_{\ttbar}$ are the signal efficiency and acceptance, respectively and $\int\mathcal{L}dt$ is the integrated luminosity.  In this section, we describe the measurement of each of these quantities.

\subsection{Event Selection and Expectation}

We select $\ttbar$ events in the lepton+jets decay channel through an inclusive lepton trigger which requires an electron (muon) with $\et>18$~GeV ($\pt>18$~GeV/$c$).  After triggering, we further require that events contain an isolated electron (muon) with $\et>20$~GeV ($\pt>20$~GeV/$c$) in the central region ($|\eta|<1.1$).  We refer to this lepton as the primary lepton, to distinguish it from the soft lepton tag.  The isolation of the primary lepton is defined as the transverse energy in the calorimeter surrounding the lepton in a cone of $\Delta R\le0.4$ $-$ but not including the lepton $\et$ itself $-$ divided by the electron (muon) $\et$ ($\pt$).  The lepton is considered isolated if the isolation is less than $0.1$.  Note that this isolation definition is different than the isolation variable, $I_\mathrm{SLT}$, which is used with the SLT$_\mathrm{e}$ algorithm.

We reject cosmic ray muons, conversion electrons, and $Z$ bosons.  Only one primary lepton is allowed to be reconstructed in the lepton+jets sample, and the flavor of that lepton must be consistent with the trigger path.  More details regarding this event selection can be found in Ref.~\cite{TopSLTmu}.  An inclusive $W$ boson sample is constructed by requiring high missing transverse energy, $\displaystyle{\not}{\et}>30$~GeV.  We suppress background events by requiring $\Ht>250$~GeV when three or more jets are present.  We define $\Ht$ as the scalar sum of the transverse energy of the primary lepton, jets, and $\displaystyle{\not}{\et}$.

In total, using events collected from February 2002 through March 2007 corresponding to an integrated luminosity of $\int\mathcal{L}dt=1.7\pm0.1$~fb$^{-1}$, we find 2196 `pretag' events with $\ge3$~jets after the event selection described.  We apply the SLT$_\mathrm{e}$ algorithm to this sample and find 120 `tag' events with $\ge 3$~jets with at least one SLT$_\mathrm{e}$, of which five have two SLT$_\mathrm{e}$ tags.  Out of 120 events, 48 have a \textsc{SecVtx} tag present, in agreement with the expected 45 such double tags.

We use \textsc{pythia} MC simulation with $m_t=175$~GeV/$c^2$ to simulate top-quark pair-production.  By default, all MC simulated samples are generated with the CTEQ5L~\cite{CTEQ} parton distribution functions (PDF), and the program \textsc{EvtGen}~\cite{EvtGen} is used to decay the particle species.  We measure $\mathcal{A}_{\ttbar}$ by counting the number of events that pass the lepton+jets event selection described above divided by the total number of events generated.  We do not restrict the decay channel at the generator level, so it is possible for some signal from other decay channels~\cite{dileptonfootnote} to be reconstructed and categorized as lepton+jets.  We then correct the acceptance with various scale factors to account for differences between simulation modeling and data.  These scale factors result from differences in modeling of the lepton identification and isolation components, as well as corrections for requirements imposed on data but not the simulation, including the trigger efficiency, the position of the primary vertex along $z$, and the quality of the lepton track.  The total acceptance for $\ttbar$ events after corrections is 6.2\%, comparable with the acceptance of other analyses in this final state~\cite{TopKinematic,TopSecVtx,TopSLTmu}.  A breakdown of the corrected acceptance by jet multiplicity and $W$ lepton type is shown in Table~\ref{table:acc}.  Scaling the acceptance by the $\ttbar$ production cross section (assumed here to be 6.7~pb) and integrated luminosity yields a total pre-tag event expectation of $716.7\pm44.4$ events, where the dominant uncertainties result from the uncertainty on the luminosity and the acceptance corrections.

Finally, we measure the efficiency to find at least one SLT$_\mathrm{e}$ tag in events that pass the event selection by applying the calorimeter requirements, tag matrix, fake matrix, and conversion efficiency scale factors to candidate tracks.  Assuming $\sigma_{\ttbar}=6.7$~pb, and $\int{\mathcal{L}dt}=1.7$~fb$^{-1}$, we expect $59.2\pm5.0$ events after tagging in the $\ge3$~jet region.  This corresponds to a per-event tagging efficiency of $\epsilon_{\ttbar}=8.3\%$.

\subsection{Background Estimation and Sample Composition}

\begin{table*}[!t]
\begin{center}
\begin{tabular}{ccc}
\hline
\hline
Process & Cross Section$\times$BF (pb) & Generator \\
\hline
$WW$ & $12.4\pm0.25$~\cite{Diboson} & \textsc{pythia} \\
$WZ$ & $3.96\pm 0.06$~\cite{Diboson} & \textsc{pythia} \\
$ZZ^*$ & $2.12\pm0.15$~\cite{Diboson} & \textsc{pythia} \\
single top ($s$-channel) & $0.29\pm0.02$~\cite{SingleTop} & \textsc{madevent}+\textsc{pythia} \\
single top ($t$-channel) & $0.66\pm0.03$~\cite{SingleTop} & \textsc{madevent}+\textsc{pythia} \\
$Z$+Jets & $308\pm51$~\cite{ZJets} & \textsc{alpgen}+\textsc{pythia} \\
Drell-Yan+Jets & $2882\pm480$~\cite{ZJets} & \textsc{alpgen}+\textsc{pythia} \\
\hline
\hline
\end{tabular}
\caption{Cross sections and generators used for the MC-simulation-derived backgrounds.  The production of single top, $Z$+jets and Drell-Yan+jets is constrained to decay (semi-)leptonically at generator level.  The cross sections for these processes are multiplied by the leptonic branching fraction.  The decay of the diboson simulation, however, remains unconstrained, and the full production cross section is quoted.}\label{table:xs}
\end{center}
\end{table*}

We consider three categories of background in the identification of $\ttbar$ events.  The first category, whose contribution is derived from MC simulation, includes the production of $WW$, $WZ$, $ZZ^*$ (where one $Z$ can be produced off-shell), single top quark production, $Z$ in association with jets, and Drell-Yan in association with jets.  These backgrounds have a small uncertainty on the production cross section or contribute sufficiently little to the total background that a large uncertainty has little effect.  For diboson production, we use \textsc{pythia} generated samples scaled by their respective theoretical cross sections to estimate their contribution to the pretag and tag samples.  The estimate for single top quark production uses a combination of \textsc{madevent}~\cite{MadEvent} for generation and \textsc{pythia} for showering, and is calculated separately for $s$- and $t$-channel processes again using the theoretical cross sections.  $Z$+jets and Drell-Yan+jets use an \textsc{alpgen}~\cite{Alpgen} and \textsc{pythia} combination, where \textsc{alpgen} is used for the generation and \textsc{pythia} is used for the showering.  The cross section is scaled to match the measured $Z$+jets cross section with an additional $1.2\pm0.2$ correction to match the measured jet multiplicity spectrum.  Table~\ref{table:xs} lists the cross sections used for each process.

\begin{figure}[htb]
\begin{center}
\includegraphics[width=3.0in]{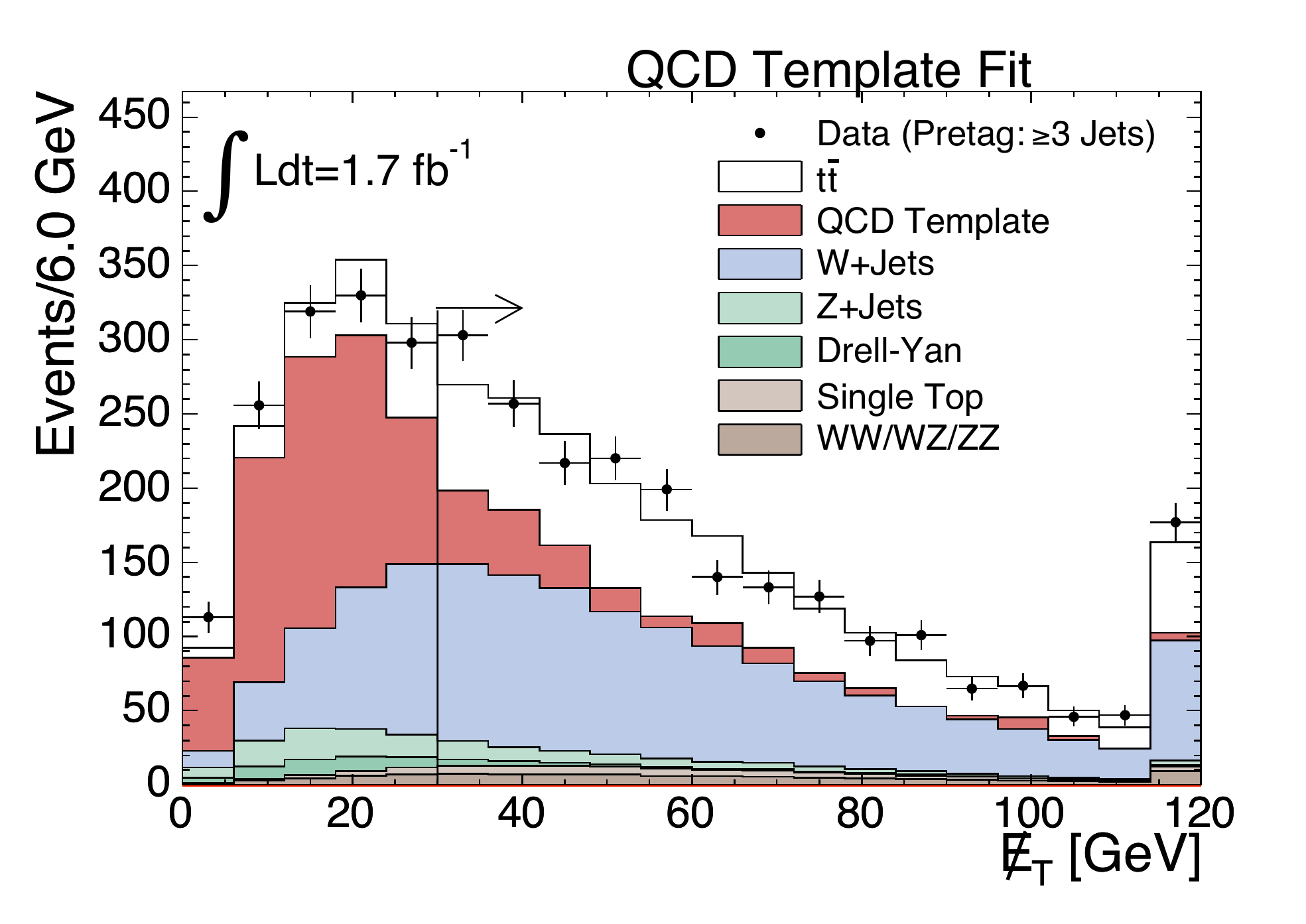}
\caption{QCD fit for pretag events with $\ge 3$ jets.  $W$+jet and QCD templates are allowed to float.}
\label{fig:qcd}
\end{center}
\end{figure}

The second category consists of background from multijet production, called QCD.  We estimate the QCD contribution by releasing the $\displaystyle{\not}{\et}$ requirement and fitting the total $\displaystyle{\not}{\et}$ distribution to templates for the backgrounds and signal.  To model the QCD $\displaystyle{\not}{\et}$ spectrum, we use two samples: a \textsc{pythia} $\bbbar$ dijet sample, and a data sample with an $\et>20$~GeV electron candidate that fails at least two electron ID requirements.  This sample is principally composed of multijet events with a similar topology to those that fake a high-$\et$ electron.  We fit for the fraction of QCD events in the sample by fixing the $\ttbar$ and MC simulation-driven background normalizations, and varying the $W$+jets and QCD template normalizations separately.  The total QCD contribution has virtually no dependence on the assumed $\ttbar$ cross section. We also include a 15\% systematic uncertainty due to the real electron contamination in the electron-like sample.  Table~\ref{table:qcd} shows the measured fits for the fraction of pretag events with $\displaystyle{\not}{\et}>30$~GeV that are due to pretag and tag QCD events, $F^{QCD}_{pre}$ and $F^{QCD}_{tag}$, respectively.  The result of the fit in the pretag region for $\ge 3$ tags is shown in Fig.~\ref{fig:qcd}.

\begin{table}[htb]
\begin{center}
\begin{tabular}{cccc}
\hline
\hline
 & 1 Jet & 2 Jets & $\ge3$~Jets \\
\hline
$F^{QCD}_{pre}$ (\%) & $3.7\pm6.0$ & $4.6\pm0.6$ & $9.2\pm1.5$ \\
$F^{QCD}_{tag}$ (\%) & $0.045\pm0.011$ & $0.10\pm0.02$ & $0.28\pm0.14$ \\
\hline
\hline
\end{tabular}
\caption{Summary of the fraction of the pretag sample due to pretag and tag QCD events for different jet multiplicies.}
\label{table:qcd}
\end{center}
\end{table}

The third category and largest background is the production of $W$ bosons in association with multiple jets.  We use a combination of simulation and data-driven techniques to measure this background.  We use \textsc{alpgen} as the generator of the $W$+multijet datasets and \textsc{pythia} for fragmentation and showering.

The $W$+jet normalization is determined by assuming that all pretag data events, not already accounted for by $\ttbar$ or by the first two background categories, must be $W$+jets.  The tag estimate is derived from the pretag estimate by assuming that the tagging efficiency measured in MC simulation for separate HF categories is accurate and only the relative amount of HF needs adjustment.  The equations below elucidate this procedure:
\begin{eqnarray}
N^{pre}_W & = & N^{pre}_{data}-N^{pre}_{MC}-N^{pre}_{QCD}-N^{pre}_{\ttbar} \label{eq:wjets}\\
N^{tag}_{W+\bbbar} & = & N^{pre}_{W}(\epsilon_{2b} F_{2b}+\epsilon_{1b}F_{1b}) \\
N^{tag}_{W+\ccbar} & = & N^{pre}_{W}(\epsilon_{2c} F_{2c}+\epsilon_{1c}F_{1c}) \\
N^{tag}_{W+{\rm LF}} & = & N^{pre}_{W}\epsilon_{0b,0c}(1-F_{2b}-F_{1b}-F_{2c}-F_{1c})
\end{eqnarray}
where $N^{tag}$ and $N^{pre}$ are the number of tag and pretag events for various signal and background components, and LF refers to light-flavor.  The tagging efficiencies, $\epsilon$, are measured in separate HF categories, where the subscript designates the number of reconstructed jets in an event identified as a $b-$ or $c-$ jet with information from the generator.  For bookkeeping purposes, the presence of a $b$-jet supersedes the presence of a $c$-jet.  The HF fractions, $F$, designate the fraction of $W$+jet events for each HF category.

\begin{table}[htb]
\begin{center}
\begin{tabular}{ccccc}
\hline
\hline
Fraction & 1 Jet & 2 Jets & 3 Jets & $\ge4$~Jets \\
\hline
$F_{1b}$ & $0.8\pm0.3$ & $1.6\pm0.6$ & $3.0\pm1.1$ & $3.7\pm1.4$ \\
$F_{2b}$ &  ---                  & $1.0\pm0.4$ & $2.2\pm0.8$ & $3.5\pm1.3$ \\
$F_{1c}$ & $5.8\pm1.6$ & $9.1\pm2.6$ & $10.2\pm3.3$ & $12.1\pm3.9$ \\
$F_{2c}$ & ---                   & $1.5\pm0.6$ & $3.4\pm1.3$ & $6.3\pm2.3$ \\
\hline
\hline
\end{tabular}
\caption{Heavy-flavor fractions multiplied by the $K$-factor for $W$+jet events.  Uncertainties are dominated by the agreement of the $K$-factor across jet bins and the $Q^2$ scale.  All numbers are shown in units of \%.}
\label{table:HFfrac}
\end{center}
\end{table}

\begin{table}[htb]
\begin{center}
\begin{tabular}{ccccc}
\hline
\hline
 & 1 Jet & 2 Jets & 3 Jets & $\ge4$~Jets \\
\hline
$\epsilon_{0b,0c}$ & $0.92\pm0.06$ & $1.89\pm0.11$ & $3.01\pm0.17$ & $4.24\pm0.24$ \\
$\epsilon_{1b}$ & $3.33\pm0.16$ & $4.39\pm0.22$ & $5.43\pm0.29$ & $6.80\pm0.36$ \\
$\epsilon_{2b}$ & --- & $6.72\pm0.33$ & $7.26\pm0.37$ & $9.55\pm0.45$ \\
$\epsilon_{1c}$ & $1.61\pm0.09$ & $2.50\pm0.14$ & $3.46\pm0.20$ & $4.78\pm0.28$ \\
$\epsilon_{2c}$ & --- & $3.11\pm0.17$ & $4.17\pm0.23$ & $5.58\pm0.30$ \\
\hline
\hline
\end{tabular}
\caption{SLT$_\mathrm{e}$ tagging efficiency for different classes of HF in $W$+jet events.  Uncertainties shown include all SLT$_\mathrm{e}$ tagging systematic uncertainties.  All numbers are shown in units of \%.}
\label{table:HFeff}
\end{center}
\end{table}

\begin{table*}[!t]
\begin{center}
\begin{tabular}{cccccc}
\hline
\hline
Process & 1 jet & 2 jets & 3 jets & 4 jets & $\ge 5$~ jets \\
\hline
Pretag & 120599 & 19695 & 1358 & 645 & 193 \\
Pretag $\ttbar$ ($\sigma=7.84$~pb) & $39.82\pm2.11$ & $211.2\pm11.2$ & $345.4\pm18.3$ & $366.6\pm19.4$ & $126.67\pm6.71 $ \\
\hline
$WW$ & $12.87\pm1.27$ & $12.36\pm1.14$ & $1.53\pm0.14$ & $0.64\pm0.06$ & $0.25\pm0.02$ \\
$WZ$ & $1.37 \pm 0.13$ & $3.04 \pm 0.26$ & $ 0.41 \pm 0.04$ & $ 0.21 \pm 0.02$ & $ 0.06 \pm 0.01 $ \\
$ZZ$ & $0.16 \pm 0.02 $ & $0.17 \pm 0.02$ & $ 0.05 \pm 0.01$ & $ 0.02 \pm 0.00$ & $ 0.01 \pm 0.00$ \\
Single Top ($s$) & $0.55 \pm 0.06$ & $ 2.31 \pm 0.23 $ & $0.46 \pm 0.05$ & $ 0.17 \pm 0.02 $ & $0.05 \pm 0.01$ \\
Single Top ($t$) & $1.88 \pm 0.17 $ & $2.67 \pm 0.25 $ & $0.36 \pm 0.03 $ & $0.09 \pm 0.01 $ & $0.01 \pm 0.00$ \\
$Z$ +Jets & $46.3 \pm 10.1$ & $ 19.52 \pm 4.02 $ & $2.44 \pm 0.44 $ & $1.09 \pm 0.20 $ & $0.28 \pm 0.05$ \\
Drell-Yan+Jets & $10.01 \pm 2.27$ & $ 6.32 \pm 1.42$ & $ 1.11 \pm 0.25 $ & $0.33 \pm 0.07 $ & $0.09 \pm 0.02$ \\
\hline
QCD & $53.9 \pm 14.1$ & $ 20.20 \pm 4.65$ & $ 3.75 \pm 1.92 $ & $1.78 \pm 0.91$ & $ 0.53 \pm 0.27$ \\
\hline
$W$+$\bbbar$ & $28.2 \pm 10.9$ & $ 22.74 \pm 8.70$ & $ 2.43 \pm 0.94$ & $ 1.04 \pm 0.43$ & $ 0.23 \pm 0.10$ \\
$W$+$\ccbar$, $W+c$ & $104.2 \pm 30.2 $ & $47.1 \pm 14.6$ & $ 3.80 \pm 1.31$ & $ 1.66 \pm 0.62 $ & $0.36 \pm 0.15$ \\
$W$+Light-Flavor & $960.7 \pm 90.8 $ & $281.0 \pm 22.9$ & $ 18.56 \pm 2.10 $ & $5.60 \pm 1.14 $ & $1.22 \pm 0.32$ \\
Total $W$+Jets & $1093 \pm 101 $ & $350.8 \pm 24.0 $ & $24.78 \pm 2.05 $ & $8.30 \pm 1.38 $ & $1.81 \pm 0.43$\\
\hline
Backgrounds & $1220.0 \pm 94.8$ & $ 417.4 \pm 25.5$ & $ 34.89 \pm 2.36 $ & $12.64 \pm 1.32 $ & $3.09 \pm 0.41$ \\
$\ttbar$ ($\sigma=7.84$~pb) & $1.41 \pm 0.10$ & $ 13.25 \pm 0.96$ & $ 26.27 \pm 1.94$ & $ 30.70 \pm 2.16$ & $ 12.41 \pm 0.86$ \\
\hline
Tags & 1312 &427 &56 &45 &19 \\
\hline
\hline
\end{tabular}
\caption{Sample composition of lepton+jet events with $\ge 1$~SLT$_\mathrm{e}$ tag corrected for the measured signal contribution.  Uncertainties include effects from luminosity, acceptance corrections, cross section uncertainties, SLT$_\mathrm{e}$ tagger modeling, $K$-factor, and the QCD fit.}\label{table:njets}
\end{center}
\end{table*}

While both the HF efficiencies and HF fractions are measured in MC simulation, the fractions are calibrated by a single, multiplicative $K$-factor, $K=1.0\pm0.4$, derived from a data/MC comparison of multijet events with HF enhanced by a \textsc{SecVtx} tag.  The systematic uncertainty is dominated by the contribution from varying the $Q^2$ of the samples and the agreement of the $K$-factor across jet multiplicities.  Phase-space overlap of jets simulated by \textsc{alpgen} and \textsc{pythia} is accounted for by allowing \textsc{alpgen} to simulate those HF jets well-separated in $\eta-\phi$ space and allowing \textsc{pythia} to simulate the rest~\cite{Sherman}.    Tables~\ref{table:HFfrac} and \ref{table:HFeff} show the measured values for the HF fractions and efficiencies, respectively.

\subsection{Measurement and Uncertainties}

\begin{figure}[htb]
\begin{center}
\includegraphics[width=3.0in]{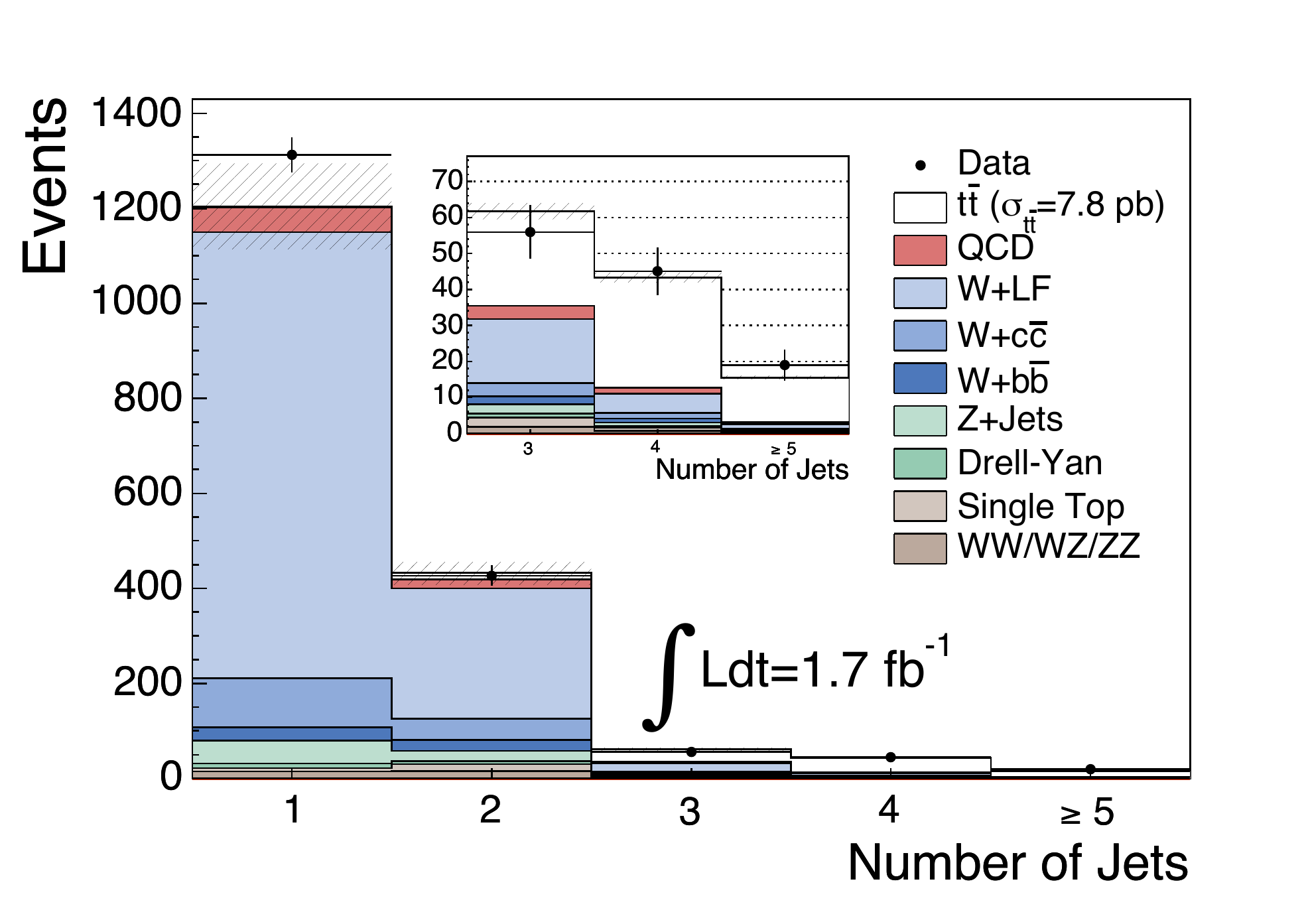}
\caption{Jet multiplicity of SLT$_\mathrm{e}$ tagged events in the lepton+jets dataset.  The embedded plot is the $\ge3$~jet subsample.  Hashed areas represent the combined systematic uncertainties, while the data shows only the statistical uncertainty.}
\label{fig:njets}
\end{center}
\end{figure}

Although the $W$+jets background depends explicitly on the assumed value of $\sigma_{\ttbar}$ (see Eq.~\ref{eq:wjets}), we can solve algebraically for the cross section, resulting in a central value of $7.8\pm2.4$~pb, where the statistical uncertainty is determined through error propagation and is verified with pseudo-experiments.  The final sample composition is shown in Table~\ref{table:njets} and is shown graphically in Fig.~\ref{fig:njets}.  This table shows the $\ttbar$ expectation for the measured cross section along with the background estimates corrected for the signal contribution.  The observed number of pretag events and the expected number of pretag $\ttbar$ events is also presented.

The combined systematic uncertainties due to the luminosity, acceptance, background cross sections, SLT$_\mathrm{e}$ tagging, $K$-factor, and QCD fit are given in the table.  Note that some of the background contributions $-$ in particular, the $W$+jets components $-$ are negatively correlated with each other, and this is reflected in the systematic uncertainties presented.  

\begin{figure*}[htb]
\begin{center}
\includegraphics[width=3.0in]{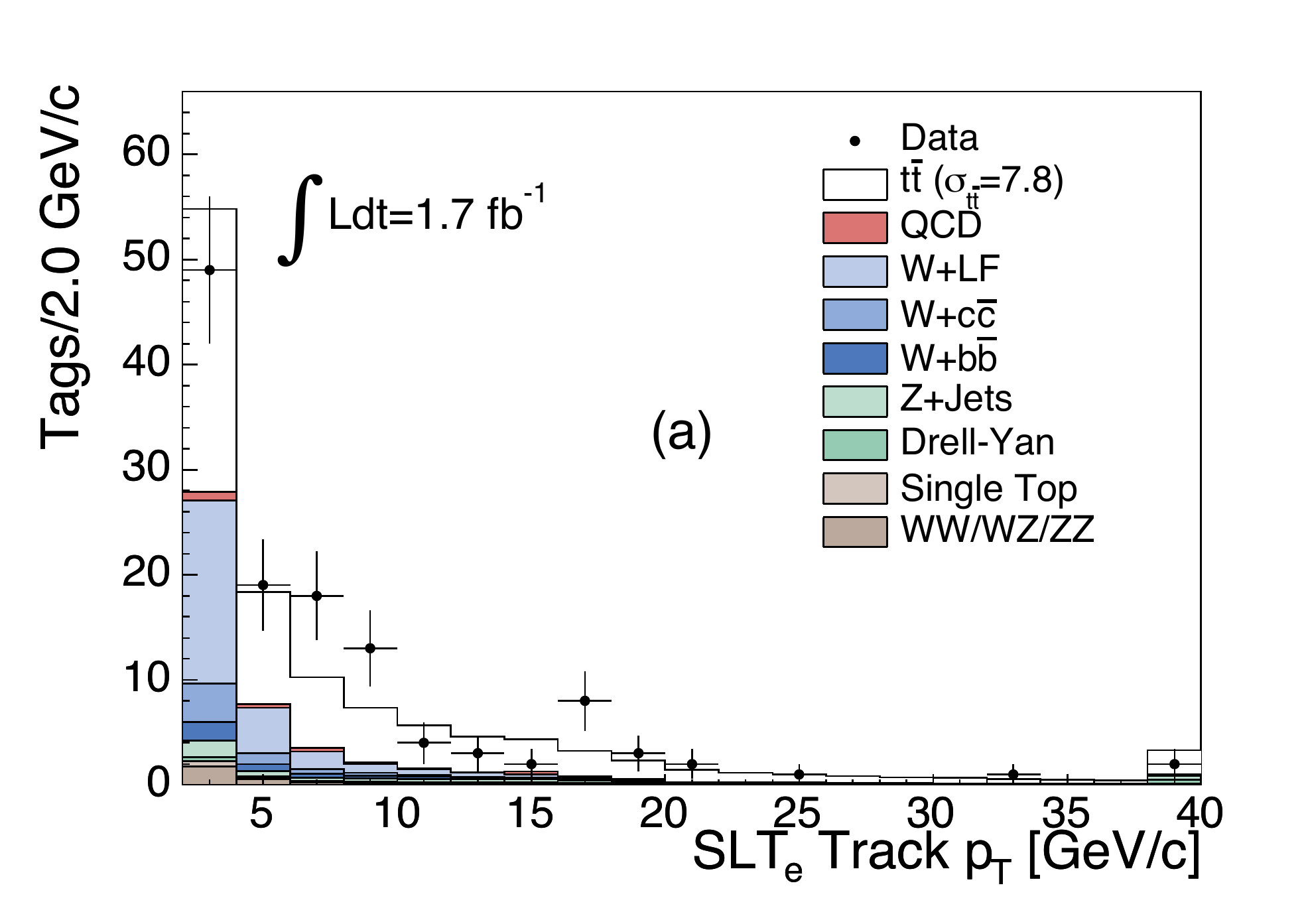}
\includegraphics[width=3.0in]{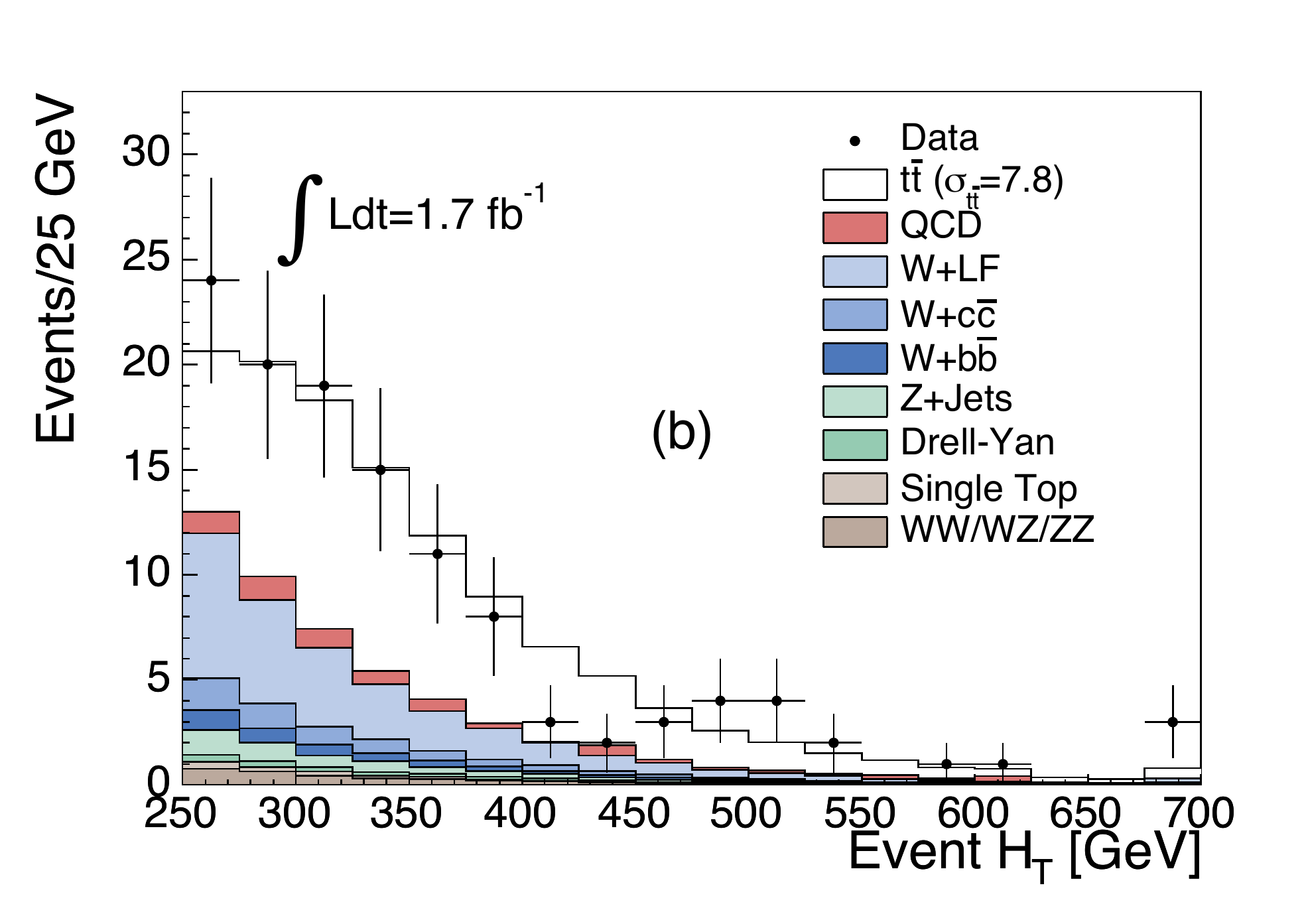}
\caption{(a) $\pt$ distribution of SLT$_\mathrm{e}$ tags in lepton+jet events with $\ge 3$ jets.  (b) $\Ht$ distribution of SLT$_\mathrm{e}$ tagged events with $\ge 3$ jets.}
\label{fig:pt}
\end{center}
\end{figure*}

Figure~\ref{fig:pt} show the SLT$_\mathrm{e}$ tag $\pt$ distribution and the event $\Ht$ distribution in the $\ge 3$~jet region.

In the previous sections, we have described systematic uncertainties related to the SLT$_\mathrm{e}$ tagger and the background estimations.  The tagger uncertainties derive from the calorimeter variable modeling, the tag- and fake-matrix predictions, the conversion (mis-)identification scale factors, and the jet environment correction from the $b$-jet tuning.  Each of the tagger uncertainties are uncorrelated because they have been derived in separate samples with distinct measurement techniques.  The background uncertainties are derived from the theoretical or experimental production cross sections, the $W$+jet HF $K$-factor, the QCD fit, and the acceptance modeling.  Here we discuss the uncertainties arising from the jet energy scale (JES) \cite{JES} and the modeling of the $\ttbar$ signal.  The systematic uncertainties are summarized in Table~\ref{table:syst}.

The effect of the JES uncertainty is calculated by adjusting the jet energy corrections that are applied to the MC simulation by $\pm 1\sigma$ and remeasuring the cross section.  The central value for the cross section is 7.2 pb with $+1\sigma$ JES and 8.5 pb with $-1\sigma$ JES, so we assign a $\pm8.6\%$ relative systematic uncertainty due to the JES.

We also determine the uncertainty from initial state radiation (ISR) and final state radiation (FSR) by remeasuring the acceptance with the \textsc{pythia} MC simulation tuned with more or less ISR and FSR.  We take the mean deviation as a systematic uncertainty.

Uncertainties related to top quark kinematic modeling and the jet fragmentation model are considered by replacing \textsc{pythia} with \textsc{Herwig} \cite{Herwig} as the event generator for the $\ttbar$ sample.  The result is a $2.2\%$ relative difference in the $\ttbar$ acceptance, which we take as a systematic uncertainty.

The uncertainty from PDFs is considered from three sources.  The first source is the difference in $\ttbar$ acceptance when the CTEQ5L PDF set is reweighted within its own uncertainties.  The second source is the difference between the CTEQ5L and an MRST98 \cite{MRST} set.  The third source is calculated by varying $\alpha_S$ within the same PDF set.  The final PDF uncertainty is calculated by taking the larger of the first two uncertainties and combining it in quadrature with the $\alpha_S$ uncertainty.  This results in a 0.9\% uncertainty on the cross section.

\begin{table}[htb]
\begin{center}
\begin{tabular}{cc}
\hline
\hline
Source & Relative Uncertainty \\
&  on $\sigma_{\ttbar}$ (\%) \\
\hline
Jet Energy Scale & 8.4 \\
QCD Fit & 5.0 \\
$K$-factor & 3.0 \\
\textsc{Herwig}/\textsc{pythia} & 2.2 \\
Acceptance Corrections & 1.6 \\
Background Cross Section & 0.6 \\
PDFs & 0.9 \\
FSR & 0.6 \\
ISR & 0.5 \\
\hline
Conversion ID Efficiency SFs & 10.7 \\
Fake Matrix & 7.8 \\
Calorimeter Modeling & 7.7 \\
Tag Matrix & 6.8 \\
Jet Environment Correction & 5.4 \\
Total Tagger Uncertainty & 17.6 \\
\hline
Total & 20.6 \\
\hline
\hline
\end{tabular}
\caption{Summary of systematic uncertainties.}
\label{table:syst}
\end{center}
\end{table}

The final result is
\begin{equation}
\sigma_{\ttbar}=7.8\pm2.4\,\mathrm{(stat)}\pm1.6\,\mathrm{(syst)}\pm0.5\,\mathrm{(lumi)}~\mathrm{pb}
\end{equation}
where we separate the luminosity uncertainty from the other systematic uncertainties.  Although we have assumed for this analysis a top quark mass of 175~GeV/$c^2$, the world average for the top quark mass is now approximately 172.4~GeV/$c^2$.  This moves the theoretical value of the cross section to approximately 7.2 pb.  The systematic uncertainty on the $\ttbar$ cross section due to the error on the top-quark mass is small, and leaves the result unchanged.

\section{Conclusions}\label{sec:conclusions}

We have performed the first measurement of the $\ttbar$ production cross section with SLT$_\mathrm{e}$ tags in Run II of the Tevatron.  This measurement, $\sigma_{\ttbar}=7.8\pm2.4\,\mathrm{(stat)}\pm1.6\,\mathrm{(syst)}\pm0.5\,$(lumi) pb, is consistent with the theoretical value~\cite{TopXS}, $\sigma_{\ttbar}=6.7\pm0.8$ pb ($m_t=175$~GeV/$c^2$), as well as the current CDF average~\cite{CDFaverage}, $\sigma_{\ttbar}=7.02\pm0.63$ pb.  While statistically limited, this measurement demonstrates the consistency of the top quark production cross section in the lepton+jets final state with soft electron $b$-tagging.  This measurement also provides an experimental basis for investigating other high-$\pt$ physics measurements with the soft electron tagging technique.\\

\begin{acknowledgments}
We thank the Fermilab staff and the technical staffs of the participating institutions for their vital contributions. This work was supported by the U.S. Department of Energy and National Science Foundation; the Italian Istituto Nazionale di Fisica Nucleare; the Ministry of Education, Culture, Sports, Science and Technology of Japan; the Natural Sciences and Engineering Research Council of Canada; the National Science Council of the Republic of China; the Swiss National Science Foundation; the A.P. Sloan Foundation; the Bundesministerium f\"ur Bildung und Forschung, Germany; the World Class University Program, the National Research Foundation of Korea; the Science and Technology Facilities Council and the Royal Society, UK; the Institut National de Physique Nucleaire et Physique des Particules/CNRS; the Russian Foundation for Basic Research; the Ministerio de Ciencia e Innovaci\'{o}n, and Programa Consolider-Ingenio 2010, Spain; the Slovak R\&D Agency; and the Academy of Finland.
\end{acknowledgments}


\begin{thebibliography}{99}
\bibitem{ptfootnote}
    We characterize the transverse energy of soft electrons by the track $\pt$, rather than the more typical calorimetric $\et$, because of the presence of the jet.
\bibitem{L00Removal}
    Although L00 is used for track reconstruction, it is only one-sided and excluded from the the conversion-finding algorithm.   The last layer of the ISL is also too far forward to be fiducial for SLT$_\mathrm{e}$ tracks.  Therefore, a total of at most six layers may be missing for any given SLT$_e$ track.
\bibitem{dileptonfootnote}
    The dilepton channel $-$ where both $W$ bosons decay leptonically $-$ contributes $2-3$\% of the total event selection as one lepton may escape identification.
\end{thebibliography}
\end{document}